\documentclass[review,12pt]{elsarticle}
\usepackage{amssymb}
\usepackage{amsthm}
\usepackage{framed} % To put a box to the nomenclature
\usepackage{amsmath,color}
\usepackage{mathrsfs}
\usepackage{graphicx}
\usepackage{epstopdf}
\usepackage{float}
\usepackage{caption}
\usepackage{subcaption}
\usepackage{bm}
\usepackage{bbm}
\usepackage{mathrsfs}
\usepackage{cleveref}
\usepackage{soul}
\usepackage{svg} % To use svg images
\usepackage{accents}
\usepackage{color,soul} %to highlight text
\usepackage{color} %To highlight references in the text
\usepackage{bm}%To apply bulky font to greek letters
\usepackage{multirow} % Vertically merged cells in tabulars
\usepackage[margin=1.78cm]{geometry}% TO PLAY WITH THE MARGINS
\biboptions{sort&compress}
\soulregister\citep7 % To allow citep to be inside of highlighted text
\soulregister\citet7 % Idem with citet
\soulregister\citealp7 % Idem with citealp
\newsavebox{\measurebox} %To create a 1 column figure with a 2 column figure on its side
\usepackage{titlesec} %To use paragraph as subsubsubsection
\usepackage[T1]{fontenc} % This is to convert the paper easily to Word
\usepackage{lmodern} % This is to convert the paper easily to Word
\input{glyphtounicode}  % This is to convert the paper easily to Word

\def\onedot{$\mathsurround0pt\ldotp$}
\def\cdddot#1{% three dots 
  \mathbin{\vcenter{\baselineskip.67ex
    \hbox{\onedot}\hbox{\onedot}\hbox{\onedot}%
  }}%
}

\journal{Theoretical and Applied Fracture Mechanics}

\makeatletter
\def\@author#1{\g@addto@macro\elsauthors{\normalsize%
    \def\baselinestretch{1}%
    \upshape\authorsep#1\unskip\textsuperscript{%
      \ifx\@fnmark\@empty\else\unskip\sep\@fnmark\let\sep=,\fi
      \ifx\@corref\@empty\else\unskip\sep\@corref\let\sep=,\fi
      }%
    \def\authorsep{\unskip,\space}%
    \global\let\@fnmark\@empty
    \global\let\@corref\@empty  %% Added
    \global\let\sep\@empty}%
    \@eadauthor={#1}
}
\makeatother

\titleformat{\paragraph}
{\normalfont\normalsize\itshape}{\theparagraph}{1em}{}
\titlespacing*{\paragraph}
{0pt}{3.25ex plus 1ex minus .2ex}{1.5ex plus .2ex}

\begin{document}

\begin{frontmatter}



\title{A general framework for decomposing the phase field fracture driving force, particularised to a Drucker-Prager failure surface}
% A general approach for decomposition strain energy density as driving force of phase field fracture method, applicaitons on the rock fracture


\author{Yousef Navidtehrani\fnref{Uniovi}}
\author{Covadonga Beteg\'{o}n \fnref{Uniovi}}
\author{Emilio Mart\'{\i}nez-Pa\~neda\corref{cor1}\fnref{IC}}
\ead{e.martinez-paneda@imperial.ac.uk}

\address[Uniovi]{Department of Construction and Manufacturing Engineering, University of Oviedo, Gij\'{o}n 33203, Spain}

\address[IC]{Department of Civil and Environmental Engineering, Imperial College London, London SW7 2AZ, UK}

\cortext[cor1]{Corresponding author.}

\begin{abstract}
Due to its computational robustness and versatility, the phase field fracture model has become the preferred tool for predicting a wide range of cracking phenomena. However, in its conventional form, its intrinsic tension-compression symmetry in damage evolution prevents its application to the modelling of compressive failures in brittle and quasi-brittle solids, such as concrete or rock materials. In this work, we present a general methodology for decomposing the phase field fracture driving force, the strain energy density, so as to reproduce asymmetrical tension-compression fracture behaviour. The generalised approach presented is particularised to the case of linear elastic solids and the Drucker-Prager failure criterion. The ability of the presented model to capture the compressive failure of brittle materials is showcased by numerically implementing the resulting strain energy split formulation and addressing four case studies of particular interest. Firstly, insight is gained into the capabilities of the model in predicting friction and dilatancy effects under shear loading. Secondly, virtual direct shear tests are conducted to assess fracture predictions under different pressure levels. Thirdly, a concrete cylinder is subjected to uniaxial and triaxial compression to investigate the influence of confinement. Finally, the localised failure of a soil slope is predicted and the results are compared with other formulations for the strain energy decomposition proposed in the literature. The results provide a good qualitative agreement with experimental observations and demonstrate the capabilities of phase field fracture methods to predict crack nucleation and growth under multi-axial loading in materials exhibiting asymmetric tension-compression fracture behaviour. \\
\end{abstract}

\begin{keyword}

Phase field fracture \sep Fracture driving force \sep Brittle fracture \sep Drucker-Prager criterion \sep Strain energy split



\end{keyword}

\end{frontmatter}



\section{Introduction}
\label{Sec:Introduction}

The application of the phase field paradigm to fracture mechanics has enabled predicting cracking phenomena of arbitrary complexity \cite{Bourdin2008,Pons2010}. These include not only hitherto complex crack trajectories but also crack branching, nucleation and merging, without \textit{ad hoc} criteria and cumbersome tracking techniques, in both two and three dimensions \cite{Borden2012,TAFM2020}. In phase field methods, the crack-solid interface is not explicitly modelled but instead smeared over a finite domain and characterised by an auxiliary phase field variable $\phi$, which takes two distinct values in each of the phases (e.g., $\phi=0$ in intact material points and $\phi=1$ inside of the crack). Hence, interfacial boundary conditions are replaced by a differential equation that describes the evolution of the phase field $\phi$. Phase field fracture methods have become the \textit{de facto} choice for modelling a wide range of cracking phenomena. New phase field formulations have been presented for ductile fracture \cite{Borden2016,Shishvan2021a}, composite materials \cite{CPB2019,Quintanas-Corominas2020a,CS2022}, shape memory alloys \cite{CMAME2021,FFEMS2022}, functionally graded materials \cite{CST2021,Kumar2021}, fatigue damage \cite{Carrara2020,CMAME2022} and hydrogen embrittlement \cite{CMAME2018,Wu2020b}, among others (see Refs. \cite{Wu2020,PTRSA2021} for an overview).\\

Most frequently, the phase field is defined to evolve in agreement with Griffith's energy balance \cite{Griffith1920} - crack growth is predicted by the exchange between elastic and fracture energies. While thermodynamically rigorous, this leads to a symmetric fracture behaviour in tension and compression, implying that crack interpenetration can occur in compressive stress states, and that the compressive strength is assumed to be equal to the tensile strength. In metals, which often fail in compression by buckling, crumbling or 45-degree shearing, this leads to nonphysical predictions of crack nucleation in compressive regions, such as the vicinity of loading pins in standardised experiments like three-point bending or compact tension. % Ref? 
For brittle and quasi brittle solids, such as concrete or geomaterials, the assumption of tension-compression symmetry is unrealistic as compressive-to-tensile strength ratios typically range between $\sigma_c/\sigma_t=2$ and $\sigma_c/\sigma_t=25$ \cite{IJRMMS2022}. In brittle materials, compressive failure takes place due to the linkage of pre-existing micro-cracks growing under local tensile stresses \cite{Sammis1986}, while tensile brittle fractures are typically due to unstable crack propagation. Thus, extending the use of phase field to the prediction of compressive failures in brittle solids requires the development of new formulations that can accommodate appropriate failure surfaces. To achieve this goal, we here present a general approach for decomposing the phase field fracture driving force, the strain energy density. We then particularise such approach to the case of a Drucker-Prager failure surface and numerically show that it can adequately capture cracking patterns in concrete and geomaterials. 

\section{The variational phase field fracture framework}
\label{Sec:Phase field fracture formulation}

We shall begin by providing a brief introduction to the variational phase field fracture formulation; the reader is referred to Ref. \cite{Bourdin2008} for a comprehensive description. Considering a body $\Omega$ with a crack surface $\Gamma$, where the displacement field $\mathbf{u}$ might be discontinuous, the energy functional can be formulated as the sum of the elastic energy stored in the cracked body and the energy required to grow the crack \cite{Francfort1998}:
\begin{equation}\label{eq:Egriffith}
    \mathcal{E} = \int_\Omega \psi \left( \bm{\varepsilon} \left( \mathbf{u} \right) \right) \, \text{d} V + \int_\Gamma G_c \, \text{d} \Gamma \, ,
\end{equation}

\noindent where $\psi$ is the elastic strain energy density, which is a function of the strain tensor $\bm{\varepsilon} \left( \mathbf{u} \right)$, and $G_c$ is a measure of the energy required to create two new surfaces, the material toughness. Equation (\ref{eq:Egriffith}) postulates Griffith's minimality principle in a global manner and its minimisation enables predicting arbitrary cracking phenomena solely as a result of the exchange between elastic and fracture energies. However, minimising Griffith's functional $\mathcal{E}$ is hindered by the unknown nature of the crack surface $\Gamma$. This can be overcome by the use of the phase field paradigm; diffusing the interface over a finite region and tracking its evolution by means of an auxiliary phase field variable $\phi$. Accordingly, Eq. (\ref{eq:Egriffith}) can be approximated by the following regularised functional:
\begin{equation}
    \mathcal{E}_\ell  = \int_\Omega g \left( \phi \right) \psi_0 \left( \bm{\varepsilon} \left( \mathbf{u} \right) \right)  \, \text{d} V + \int_{V} G_c \gamma \left( \phi,\nabla \phi, \ell \right) \, \text{d} V  \, ,
    \label{eq:E_ell}
\end{equation}

\noindent where $\psi_0$ denotes the elastic strain energy density of the undamaged solid, $g (\phi)$ is a degradation function to reduce the stiffness of the solid with increasing damage, and $\gamma \left( \phi,\nabla \phi, \ell \right)$ is the so-called crack density function. For simplicity, and without loss of generality, we adopt the constitutive choices of the so-called conventional or \texttt{AT2} phase field model \cite{Bourdin2000}, such that
\begin{equation}
   g \left( \phi \right) =  \left( 1-\phi \right)^2 \,\,\,\,\,\,\, \text{and} \,\,\,\,\,\,\,  \gamma (\phi,\nabla, \ell \phi) = \frac{1}{2\ell}\phi^2 + \frac{\ell}{2} \lvert\nabla \phi\rvert^{2} \,
   \label{eq:ConsPF}
\end{equation}

\noindent where $\ell$ is the phase field length scale, inherently arising due to the non-local nature of the model. The strong form of the balance equations can be derived by taking the first variation of $\mathcal{E}_\ell$ with respect to the primal kinematic variables ($\mathbf{u}$, $\phi$) and making use of Gauss' divergence theorem, rendering
\begin{align}\label{eq:strongForm}
\nabla \cdot \left[(1-\phi)^2  \boldsymbol{\sigma}_0 \right] &= \boldsymbol{0}   \hspace{3mm} \rm{in}  \hspace{3mm} \Omega \nonumber \\ 
G_{c}  \left( \dfrac{\phi}{\ell}  - \ell \nabla^2 \phi \right) - 2(1-\phi) \, \psi_0  &= 0 \hspace{3mm} \rm{in} \hspace{3mm} \Omega  
\end{align}

\noindent where $\boldsymbol{\sigma}_0$ is the undamaged stress tensor. As seen in (\ref{eq:strongForm})b, the evolution of the phase field is governed by the (undamaged) elastic strain energy density which, for linear elastic isotropic solids, is given by
\begin{equation}\label{eq:psi0}
    \psi_0 = \frac{1}{2} \lambda \text{tr} \left( \bm{\varepsilon} \right)^2 + \mu \, \bm{\varepsilon} : \bm{\varepsilon} \, ,
\end{equation}

\noindent where $\lambda$ and $\mu$ are the Lamé coefficients. It follows that the phase field is insensitive to the compressive or tensile nature of the mechanical fields (tension-compression symmetry in damage evolution). To enforce a distinction between tension and compression behaviour, several formulations have been proposed. Initially, the motivation was the need to avoid crack interpenetration and achieve the resistance to cracking under compression observed in some materials such as metals. Examples of strain energy decompositions formulated with this objective include the volumetric-deviatoric split by Amor \textit{et al.} \cite{Amor2009}, the spectral decomposition by Miehe and co-workers \cite{Miehe2010a}, and the purely tensile splits (so-called 'no-tension' models) of Freddi and Royer-Carfagni \cite{Freddi2010,Freddi2011} and Lo \textit{et al.} \cite{Lo2019}. On the other hand, rising interest in using phase field methods to model fracture in concrete and geomaterials has led to the development of driving force definitions that accommodate non-symmetric failure surfaces \cite{Choo2018}. Zhou \textit{et al.} \cite{Zhou2019b} and Wang \textit{et al.} \cite{Wang2019b} developed new driving force formulations based on Mohr-Coulomb theory. And very recently, de Lorenzis and Maurini \cite{Lorenzis2021} presented an analytical study where the strian energy split was defined based on a Drucker-Prager failure surface. The majority of these works adopt the following structure. The elastic strain energy density is decomposed into two parts: (i) a part affected by damage, $\psi_d$, and (ii) a stored residual elastic part $\psi_s$, which is independent of the damage variable and thus not susceptible to dissipation. Accordingly,
\begin{equation}\label{eq:Split}
    \psi_0 \left( \bm{\varepsilon} \right) = \psi_d \left( \bm{\varepsilon} \right) + \psi_s \left( \bm{\varepsilon} \right) \, , \,\,\,\,\,\, \text{and} \,\,\,\,\,\, \psi \left( \bm{\varepsilon}, \phi \right) = g \left( \phi \right) \psi_d \left( \bm{\varepsilon} \right) + \psi_s \left( \bm{\varepsilon} \right) ,
\end{equation}
%\begin{equation}
%    \psi \left( \bm{\varepsilon}, \phi \right) = g \left( \phi \right) \psi_d \left( \bm{\varepsilon} \right) + \psi_s \left( \bm{\varepsilon} \right)
%\end{equation}
\noindent which necessarily implies,
\begin{equation}
    \psi \left( \bm{\varepsilon}, \phi \right) = g \left( \phi \right) \psi_0 \left( \bm{\varepsilon} \right) + \left( 1 - g \left( \phi \right) \right) \psi_s \left( \bm{\varepsilon} \right) \, .
\end{equation}

And this decomposition of the strain energy density gives rise to an analogous decomposition of the Cauchy stress tensor, such that
\begin{equation}
    \sigma \left( \bm{\varepsilon} , \phi \right) = g \left( \phi \right) \frac{\partial \psi_d \left( \bm{\varepsilon} \right)}{\partial \bm{\varepsilon}} + \frac{\partial \psi_s \left( \bm{\varepsilon} \right)}{\partial \bm{\varepsilon}} =g \left( \phi \right) \bm{\sigma}^d+ \bm{\sigma}^s \, .
\end{equation}
\noindent where $\bm{\sigma}^d$ and $\bm{\sigma}^s$ respectively denote the damaged and non-degraded parts of the Cauchy stress tensor.\\

The aim of this work is to present a generalised approach to identify $\psi_s \left( \bm{\varepsilon} \right)$ (and subsequently $\psi_d \left( \bm{\varepsilon} \right)$) as a function of the failure surface and the constitutive behaviour of the pristine material. This is presented below, in Section \ref{Sec:General approach}, where the framework is exemplified with a Drucker-Prager \cite{Drucker1952} failure surface. 
%\textcolor{blue}{We emphasise that, in this work, undamaged material points are characterised by $\phi=0$, the nucleation or onset of damage occurs when $\phi>0$ and the nucleation of a crack (i.e., the formation of a new singularity) takes place when the phase field first reaches $\phi=1$ in a previously undamaged region.}

\section{A general approach for decomposing the strain energy density based on failure criteria}
\label{Sec:General approach}

We proceed to present a general approach for decomposing the strain energy density so as to incorporate any arbitrary failure criterion in the phase field fracture method. As the strain energy density is the driving force for fracture, a suitable choice of strain energy decomposition can enable reproducing the desired failure surface. Such a choice must satisfy the failure criterion assumed while recovering the constitutive behaviour of the pristine material. Here, for simplicity, we choose to focus on solids exhibiting linear elastic behaviour in the undamaged state. However, the framework is general and can be extended to other constitutive responses, such as hyperelasticity. We shall first derive the partial differential equation (PDE) that characterises the possible solutions for the non-dissipative stored strain energy density $\psi_s$ in linear elastic solids. Then, we consider the failure envelope function that provides the constraint required to obtain a solution to this PDE. The process is exemplified with a Drucker-Prager failure surface, and the section concludes with brief details of the numerical implementation.\\ 

\begin{figure}[H]
    \centering
    \begin{subfigure}[b]{0.19\textwidth} 
    \includegraphics[width=\textwidth]{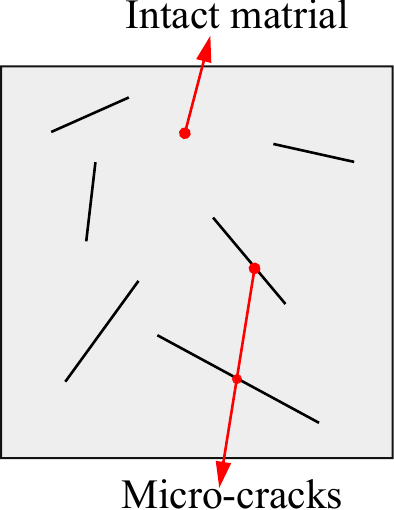} 
    \caption{}
    \label{fig:RVE-a}
    \end{subfigure} \hspace{20 mm}
    \begin{subfigure}[b]{0.25\textwidth} 
    \includegraphics[width=\textwidth]{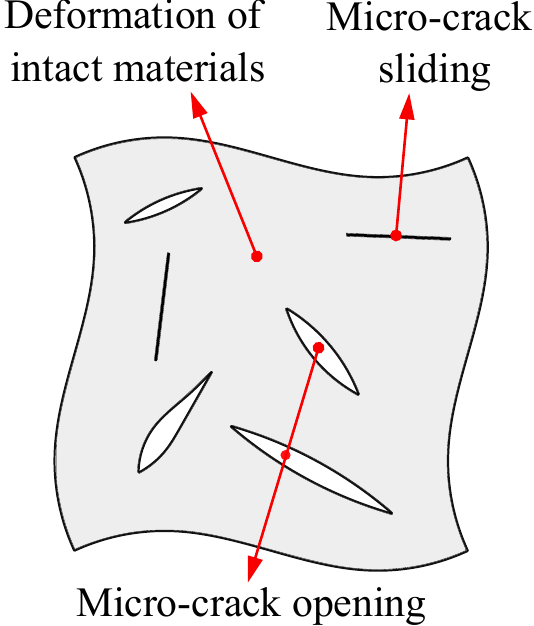}
    \caption{}
    \label{fig:RVE-b}
    \end{subfigure}
    \caption{Meso-scale Representative Volume Element (RVE) of a damaged solid, showing regions of micro-cracks and intact material in the: (a) undeformed, and (b) deformed states, with the latter emphasising the effect of micro-crack opening and sliding.}
    \label{fig:RVE}
\end{figure}

As in Ref. \cite{Freddi2010}, the \emph{Theory of Structured Deformations} \cite{DelPiero1993} is applied to a damaged continuum solid. We confine our attention to infinitesimal deformations, such that the total strain tensor can be estimated from the displacement vector as,
\begin{equation}
    \bm{\varepsilon} = \frac{1}{2} \left( \nabla \mathbf{u}^T + \nabla \mathbf{u}  \right)
\end{equation}
A Representative Volume Element (RVE) can be defined, see Fig. \ref{fig:RVE}, such that the meso-scale representation of the material involves regions of intact material and micro-cracks. In this context, the phase field $\phi$ is akin to a damage variable, and describes the integrity of the RVE (the extent of dominance of intact and cracked regions, within the two limiting cases of $\phi=0$ and $\phi=1$). The macroscopic deformation is then the sum of two contributions: an elastic straining of the intact material regions, and the opening and sliding of micro-cracks, that can coalescence into macroscopic cracks. Accordingly, 
\begin{equation}
    \boldsymbol{\varepsilon}=\boldsymbol{\varepsilon}^e+\boldsymbol{\varepsilon}^d \, ,
\end{equation}

\noindent where $\bm{\varepsilon}^e$ are the elastic (recoverable) strains due to the deformation of the undamaged structure, while $\bm{\varepsilon}^d$ denotes the inelastic strains associated with microscopic damage mechanisms. 

The elastic strain tensor $\boldsymbol{\varepsilon}^e$ is related to the Cauchy stress tensor through the inverse of the elastic stiffness matrix $\boldsymbol{\varepsilon}^e=\left(\mathcal{C}_0\right)^{-1} \boldsymbol\sigma$ and, if $\boldsymbol{\varepsilon}^e$ and $\boldsymbol{\varepsilon}^d$ are orthogonal, the stored and damaged strain energy densities of effective configuration (see Section \ref{Sec:Phase field fracture formulation}) can be estimated as,
\begin{equation}
    \psi_s=\frac{1}{2} \boldsymbol{\varepsilon}^e C_0 \boldsymbol{\varepsilon}^e \,\,\,\,\,\,\,\,\, \text{and} \,\,\,\,\,\,\,\,\, \psi_d=\frac{1}{2} \boldsymbol{\varepsilon}^d C_0 \boldsymbol{\varepsilon}^d
\end{equation}

\noindent with the total strain energy density $\psi$ being computed from $\psi_s$ and $\psi_d$ using Eq. (\ref{eq:Split}). Now, let us consider the strain energy density of pristine material as a function of the effective stress invariants ($I_1(\boldsymbol{\sigma}_0),J_2(\boldsymbol{\sigma}_0)$);
\begin{equation}
\psi_0(\boldsymbol{\varepsilon})=\frac{1}{18 K} I_1^2 (\boldsymbol\sigma_0(\boldsymbol{\varepsilon}))+\frac{1}{2 \mu} J_2 (\boldsymbol\sigma_0(\boldsymbol{\varepsilon})) \, ,
\label{Eq:EffectiveElasticStrainEnergy}
\end{equation}
\noindent where $K$ is the bulk modulus, $\mu$ is the shear modulus, $I_1$ is the first invariant of a tensor, and $J_2$ is the second invariant of the deviatoric part of a tensor. Eq. (\ref{Eq:EffectiveElasticStrainEnergy}) holds for any linear elastic isotropic solid. The stiffness and material behaviour associated with the non-degraded strain energy density $\psi^s$ and stress $\bm{\sigma}^s$ corresponds to that of intact material and, accordingly, 
\begin{equation}
\psi_s=\frac{1}{18 K} I_1^2 (\boldsymbol\sigma^s)+\frac{1}{2 \mu} J_2 (\boldsymbol\sigma^s) \, .
\label{Eq:ElasticStrainEnergy}
\end{equation}
Then, for any choice of $\psi(I_1(\bm{\varepsilon}),J_2(\bm{\varepsilon}))$, it is possible to describe the relation between the invariants of strain and stress as follows (see \ref{App:I1J2Proof}):
\begin{equation}
 I_1 (\boldsymbol\sigma (\boldsymbol\varepsilon))=3 \frac{\partial \psi(\boldsymbol\varepsilon)}{\partial I_1(\boldsymbol\varepsilon)}, \quad J_2 (\boldsymbol\sigma (\boldsymbol\varepsilon))=J_2(\boldsymbol\varepsilon) \left(\frac{\partial \psi(\boldsymbol\varepsilon)}{\partial J_2(\boldsymbol\varepsilon)} \right)^2
\label{Eq:strain stress relation}
\end{equation}

By substituting Eq. (\ref{Eq:strain stress relation}) into Eq. (\ref{Eq:ElasticStrainEnergy}), one can obtain the PDE for the stored strain energy density,
\begin{equation}
\psi_s=\frac{1}{2 K} \left(\frac{\partial \psi_s}{\partial I_1(\boldsymbol\varepsilon)}\right)^2+\frac{J_2(\boldsymbol\varepsilon)}{2 \mu} \left(\frac{\partial \psi_s}{\partial J_2(\boldsymbol\varepsilon)} \right)^2
\label{Eq:PDE strain energy cracked}
\end{equation}

Upon the appropriate constraints and boundary conditions, one can solve the PDE (\ref{Eq:PDE strain energy  cracked}) to obtain the non-dissipative stored part of the strain energy density for any level of material damage. The additional constraint needed comes from the definition of the failure criterion under consideration. Any arbitrary failure envelope can be defined in terms of the stress invariants for the fully damaged state. For illustration, let us consider a failure surface defined in terms of $I_1$ and $J_2$; i.e., $f\left( I_1 (\boldsymbol\sigma^f),J_2 (\boldsymbol\sigma^f) \right)=0$, where $\boldsymbol\sigma^f=\boldsymbol\sigma(\bm{\varepsilon},\phi=1)$. Accordingly, considering Eq. (\ref{Eq:strain stress relation}), the following failure envelope function can be defined:
\begin{equation}
    f\left(\frac{\partial \psi_s(\boldsymbol\varepsilon)}{\partial I_1(\boldsymbol\varepsilon)},\frac{\partial \psi_s(\boldsymbol\varepsilon)}{\partial J_2(\boldsymbol\varepsilon)} \right)=0 \,
    \label{Eq:PDE of criteria}
\end{equation}

\noindent and $\psi_s$ can be found from the common solution to Eqs. (\ref{Eq:PDE strain energy  cracked}) and (\ref{Eq:PDE of criteria}) upon the application of appropriate boundary conditions. This is showcased below for a Drucker-Prager failure envelope. 

\subsection{Particularisation to the Drucker-Prager failure surface}

Drucker-Prager’s failure criterion was developed for pressure-dependent materials like rock, concrete, foams and polymers. In terms of invariants of stress, the Drucker-Prager criterion is expressed as follows,
\begin{equation}
    \sqrt{J_2(\boldsymbol\sigma)}=A+B I_1 (\boldsymbol\sigma) \, ,
    \label{Eq:Drucker-prager}
\end{equation}

\noindent where $A$ and $B$ are a function of the uniaxial tensile ($\sigma_t$) and compressive ($\sigma_c$) strengths, such that
\begin{equation}
A=\frac{2}{\sqrt{3}}\left(\frac{\sigma_{c} \sigma_{t}}{\sigma_{c}+\sigma_{t}}\right) \,; \, \,  \,\,\,\,\,\ B=\frac{1}{\sqrt{3}}\left(\frac{\sigma_{t}-\sigma_{c}}{\sigma_{c}+\sigma_{t}}\right) \, .
\end{equation}

A material point sitting inside the Drucker-Prager failure envelope can be assumed to behave in a linear elastic manner, with damage-driven non-linear behaviour being triggered when the stress state reaches the failure surface. Assuming that the same degradation function $g(\phi)$ applies to the tensile and compressive strengths, then the sensitivity of the parameters $A$ and $B$ to the phase field variable is characterised by, 
\begin{equation}
\begin{aligned}
& A(\phi)=\frac{2}{\sqrt{3}}\left(\frac{g(\phi)\sigma_{c} g(\phi)\sigma_{t}}{g(\phi)\sigma_{c}+g(\phi)\sigma_{t}}\right)=g(\phi) \frac{2}{\sqrt{3}}\left(\frac{\sigma_{c} \sigma_{t}}{\sigma_{c}+\sigma_{t}}\right)=g(\phi) A(\phi=0) \\ &  B(\phi)=\frac{1}{\sqrt{3}}\left(\frac{g(\phi)\sigma_{t}-g(\phi)\sigma_{c}}{g(\phi)\sigma_{c}+g(\phi)\sigma_{t}}\right)=\frac{1}{\sqrt{3}}\left(\frac{\sigma_{t}-\sigma_{c}}{\sigma_{c}+\sigma_{t}}\right)=B(\phi=0)
\end{aligned}
\label{Eq:Drucker-prager param phi}
\end{equation}

Accordingly, for the fully damaged state ($\phi=1$), the Drucker-Prager parameters read,
\begin{equation}
A(\phi=1)=0 \, ; \,\,\,\,\,\,\,\,\ \quad B(\phi=1)=B(\phi=0) \, .
\end{equation}

\noindent I.e., $A$ is degraded as the phase field evolves, while the parameter $B$ is insensitive to the damage state. This can be physically interpreted through the cohesion parameter $c$ and the friction angle $\theta$ of Mohr-Coulomb's criterion, and their relationship with Drucker-Prager's coefficients:
\begin{equation}\label{eq:ABctheta}
A \left( \theta, c \right) = \frac{6 c \cos \theta}{\sqrt{3}(3+\sin \theta)}\, ; \,\,\,\,\,\ B \left( \theta \right) = \frac{2 \sin \theta}{\sqrt{3}(3+\sin \theta)} \, .
\end{equation}

As seen in Eq. (\ref{eq:ABctheta}), $B$ is only a function of the friction angle, while $A$ is also a function of $c$, exhibiting a linear relationship with the cohesion parameter. Since damage translates into a loss of cohesion, both $A$ and $c$ degrade with evolving damage, and eventually vanish in fully cracked state.\\

\begin{figure}[H]
    \centering
    \includegraphics[width=1\textwidth]{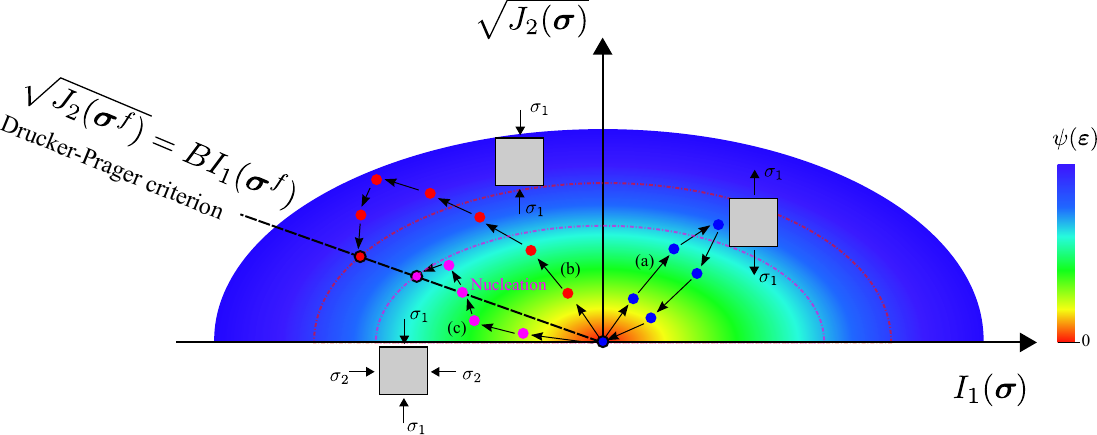}
    \caption{Stress states in the $\left(I_1 (\boldsymbol\sigma),\sqrt{J_2(\boldsymbol\sigma)}\right)$. Three loading paths have been schematically incorporated to showcase the three potential scenarios discussed in Eqs. (\ref{eq:psi_s3regimes}) and (\ref{eq:psi_d3regimes}), and colour contours denote the magnitude of the total strain energy (increasing as we move away from the origin). Circles with an outer black domain denote fully damaged states ($\phi=1$).}
    \label{fig:I1J2}
\end{figure}

In addition, consistent with Eq. (\ref{Eq:Drucker-prager}), the stress state in the fully damaged configuration satisfies,
\begin{equation}
    \sqrt{J_2(\boldsymbol\sigma^f)}=B I_1 (\boldsymbol\sigma^f) \,
    \label{Eq:Drucker-prager1} \, ,
\end{equation}

\noindent as the stress state goes back to the failure envelope for $\phi=1$ (see Fig. \ref{fig:I1J2}).\\

As discussed above, our general approach requires a function describing the failure condition in terms of the strain energy density and the strains - see Eq. (\ref{Eq:PDE of criteria}). This can be achieved by combining Eqs. (\ref{Eq:strain stress relation}) and (\ref{Eq:Drucker-prager1}), reaching
\begin{equation}
    f\left(\frac{\partial \psi_s(\boldsymbol\varepsilon)}{\partial I_1(\boldsymbol\varepsilon)},\frac{\partial \psi_s(\boldsymbol\varepsilon)}{\partial J_2(\boldsymbol\varepsilon)} \right)=
    \sqrt{J_2(\boldsymbol\varepsilon)}\frac{\partial \psi_s(\boldsymbol\varepsilon)}{\partial J_2(\boldsymbol\varepsilon)}-3 B \frac{\partial \psi_s(\boldsymbol\varepsilon)}{\partial I_1(\boldsymbol\varepsilon)}=0
    \label{Eq:PDE Drucker-Prager}
\end{equation}

An isotropic linear elastic material must satisfy Eq. (\ref{Eq:PDE strain energy  cracked}) and, if obeying the Drucker-Prager failure criterion, also Eq. (\ref{Eq:PDE Drucker-Prager}). Hence, the common solution to these two PDEs will give us the stored (elastic) strain energy density $\psi_s$. Let us obtain this common solution by first finding the general solution of Eq. (\ref{Eq:PDE Drucker-Prager}), which is of the form
\begin{equation}
    \psi_s=a_1 \left(I_1(\boldsymbol\varepsilon)+6 B \sqrt{J_2(\boldsymbol\varepsilon)} \right)^2+ a_2
    \label{Eq:Drucker-Prager Strain Energy}
\end{equation}

\noindent where $a_1$ and $a_2$ are unknowns. These can be estimated by applying suitable boundary conditions and substituting the general solution into the second PDE. Hence, considering the boundary condition $\psi_s (I_1(\boldsymbol\varepsilon)=0,J_2(\boldsymbol\varepsilon)=0)=0$, one finds that $a_2=0$. Then, the remaining unknown is obtained by deriving Eq. (\ref{Eq:Drucker-Prager Strain Energy}) with respect to $I_1(\boldsymbol\varepsilon)$ and $J_2(\boldsymbol\varepsilon)$ and substituting into Eq. (\ref{Eq:PDE strain energy cracked}), rendering
\begin{equation}
a_1=\frac{K \mu}{18 B^2 K + 2 \mu} \, .
\end{equation}

Accordingly, upon substitution in Eq. (\ref{Eq:Drucker-Prager Strain Energy}), the stored (elastic) strain energy density associated with the Drucker-Prager failure envelope is found to be:
\begin{equation}
    \psi_s=\frac{K \mu}{18 B^2 K + 2 \mu} \left(I_1(\boldsymbol\varepsilon)+6 B \sqrt{J_2(\boldsymbol\varepsilon)} \right)^2
    \label{Eq:Drucker-Prager Strain Energy1}
\end{equation}

However, one should note that Eq. (\ref{Eq:Drucker-Prager Strain Energy1}) is only valid for stress states that are above the failure envelope. Three potential scenarios exist: (1) the first invariant of stress is positive, $I_1(\boldsymbol\sigma)>0$; (2) the stress state is above the failure criterion, $\sqrt{J_2(\boldsymbol\sigma)} \geq B I_1 (\boldsymbol\sigma)$; and (3) the stress state is below the failure criterion, $\sqrt{J_2(\boldsymbol\sigma)}<B I_1 (\boldsymbol\sigma)$. With scenarios (2) and (3) being only relevant when $I_1(\boldsymbol\sigma)<0$. We then proceed to generalise Eq. (\ref{Eq:Drucker-Prager Strain Energy1}) to encompass those three regimes (see \ref{App:3regimes}), such that
\begin{equation}
\psi_s=
\begin{cases}
0 & \text{for} \quad -6B\sqrt{J_2 (\boldsymbol\varepsilon)} < I_1 (\boldsymbol\varepsilon) \\
\frac{K \mu}{18 B^2 K + 2 \mu} \left(I_1(\boldsymbol\varepsilon)+6 B \sqrt{J_2(\boldsymbol\varepsilon)} \right)^2 & \text{for} \quad -6B\sqrt{J_2 (\boldsymbol\varepsilon)} \geq I_1 (\boldsymbol\varepsilon) \,\, \& \,\,  2 \mu \sqrt{J_2 (\boldsymbol\varepsilon)} \geq 3 B K I_1 (\boldsymbol\varepsilon) \\
\frac{1}{2}K I_1^2 (\boldsymbol\varepsilon)+2 \mu J_2 (\boldsymbol\varepsilon) & \text{for} \quad 2 \mu \sqrt{J_2 (\boldsymbol\varepsilon)} < 3 B K I_1 (\boldsymbol\varepsilon) 
\end{cases}
\label{eq:psi_s3regimes}
\end{equation}

\noindent And the damaged part of the strain energy density can be readily estimated using Eq. (\ref{eq:Split}), rendering
\begin{equation}
\psi_d=
\begin{cases}
\frac{1}{2}K I_1^2 (\boldsymbol\varepsilon)+2 \mu J_2 (\boldsymbol\varepsilon) & \text{for} \quad -6B\sqrt{J_2 (\boldsymbol\varepsilon)} < I_1 (\boldsymbol\varepsilon) \\
\frac{1}{18 B^2 K + 2 \mu} \left(-3BKI_1(\boldsymbol\varepsilon)+2 \mu \sqrt{J_2(\boldsymbol\varepsilon)} \right)^2 & \text{for} \quad -6B\sqrt{J_2 (\boldsymbol\varepsilon)} \geq I_1 (\boldsymbol\varepsilon) \,\, \& \,\,  2 \mu \sqrt{J_2 (\boldsymbol\varepsilon)} \geq 3 B K I_1 (\boldsymbol\varepsilon)\\
0 & \text{for} \quad 2 \mu \sqrt{J_2 (\boldsymbol\varepsilon)} < 3 B K I_1 (\boldsymbol\varepsilon) 
\end{cases}
\label{eq:psi_d3regimes}
\end{equation}

The different stress states are illustrated in Fig. \ref{fig:I1J2} in terms of their location in the $\left(I_1 (\bm{\sigma}), \sqrt{J_2 (\bm{\sigma})}\right)$ space, where the colour contours denote the magnitude of the total strain energy (increasing as we move away from the origin). The loading path illustrated with blue dots, path (a), illustrates the case where the first invariant of stress is positive $I_1(\boldsymbol\sigma)>0$. In such a scenario, the failure process is driven by $\psi_d$, with the fully damage state achieved by returning to the origin (where the loading path intersects the Drucker-Prager failure criterion). In regards to the stress states on the left side of the figure ($I_1(\boldsymbol\sigma)<0$), their behaviour is differentiated by their location relative to the Drucker-Prager criterion, which is represented by the $\sqrt{J_2(\boldsymbol\sigma)} = B I_1 (\boldsymbol\sigma)$ line. Thus, the red loading path (b) is above the Drucker-Prager criterion and both $\psi_s$ and $\psi_c$ are active, see Eqs. (\ref{eq:psi_s3regimes})b and (\ref{eq:psi_d3regimes})b. Eventually, the loading path intersects again the $\sqrt{J_2(\boldsymbol\sigma)} = B I_1 (\boldsymbol\sigma)$ line, reaching the fully damaged state and the associated residual strain energy density $\psi_s$. Finally, loading paths within the $I_1(\boldsymbol\sigma)<0$ domain can also lie below the failure criterion, as showcased by the purple circles, path (c). In this case, $\psi_d=0$, see Eq. (\ref{eq:psi_d3regimes})c, and consequently $\phi=0$. As shown in Fig. \ref{fig:I1J2}, changes in stress state associated with the loading path might lead to an intersection with the Drucker-Prager failure line, in what would constitute a micro-fracturing nucleation event ($\phi>0$). Subsequently, final rupture ($\phi=1$) would be attained when the loading path intersects again with the failure line, rendering a residual strain energy density $\psi_s$.\\

This phase field fracture formulation built upon Drucker-Prager's failure criterion is numerically implemented using the finite element method. Retaining unconditional stability, we solve in a monolithic fashion the coupled system of equations that results from restating the local force balances,
\begin{align}
\nabla \cdot \left[\left( 1 - \phi \right)^2 \frac{\partial \psi_d \left( \bm{\varepsilon} \right)}{\partial \bm{\varepsilon}} + \frac{\partial \psi_s \left( \bm{\varepsilon} \right)}{\partial \bm{\varepsilon}} \right] &= \boldsymbol{0}   \hspace{3mm} \rm{in}  \hspace{3mm} \Omega \nonumber \\ 
G_{c}  \left( \dfrac{\phi}{\ell}  - \ell \nabla^2 \phi \right) - 2 (1 - \phi)  \mathcal{H}  &= 0 \hspace{3mm} \rm{in} \hspace{3mm} \Omega  
\end{align}

\noindent into their weak form. Here, $\mathcal{H}=\text{max} \, \psi_d (t)$ is a history field introduced to enforce damage irreversibility \cite{Miehe2010a}. As described in \ref{App:FEM}, we take advantage of the analogy between the phase field evolution law and the heat transfer equation to implement the model into the finite element package \texttt{ABAQUS} using solely a user-material subroutine (\texttt{UMAT}) (see Refs. \cite{Materials2021,AES2021}).

\section{Representative results}

Now, we shall illustrate the potential of enriching the phase field fracture description with a failure envelope of our choice. Specifically. through numerical examples, we will showcase how a formulation based on the Drucker-Prager failure criterion can capture the compressive failure of brittle materials such as concrete or geomaterials, along with capturing frictional behaviour and the dilatancy effect. Firstly, in Section \ref{Sec:RSingleElement}, we gain insight into the material behaviour resulting from the Drucker-Prager strain energy split adopted by investigating the response of a single element undergoing shear. Secondly, numerical experiments using the Direct Shear Test (DST) configuration are conducted in Section \ref{Sec:RDST}. The goal is to investigate the fracture predictions obtained under the conditions relevant to the determination of the failure properties of frictional materials. The third case study, shown in Section \ref{Sec:Rcompressionconcrete}, involves conducting virtual uniaxial and triaxial compression tests on concrete, so as to investigate the confinement effect. Finally, in Section \ref{Sec:Rsoil}, the predictions obtained from three strain energy splits are compared in the modelling of the localised failure of a soil slope. Our finite element calculations extend the very recent analytical study by de Lorenzis and Maurini \cite{Lorenzis2021}, where a Drucker-Prager failure surface was also adopted. 

\subsection{Single element under shear deformation}
\label{Sec:RSingleElement}

\begin{figure}[H]
    \centering
    \begin{subfigure}[t]{0.4\textwidth}
    \includegraphics[width=\textwidth]{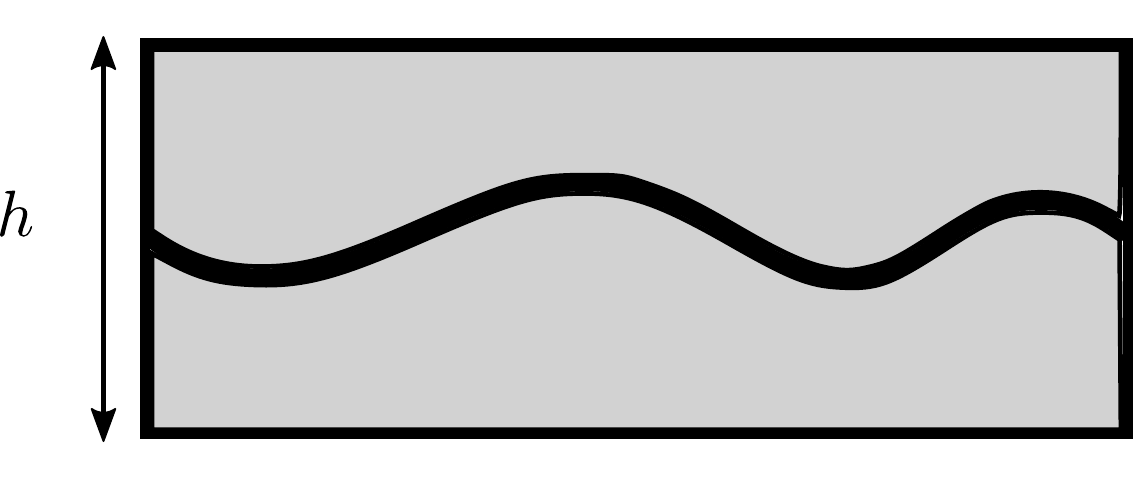}
    \caption{}
    \label{}
    \end{subfigure} \hspace{1 cm}
    \begin{subfigure}[t]{0.47\textwidth} 
    \includegraphics[width=\textwidth]{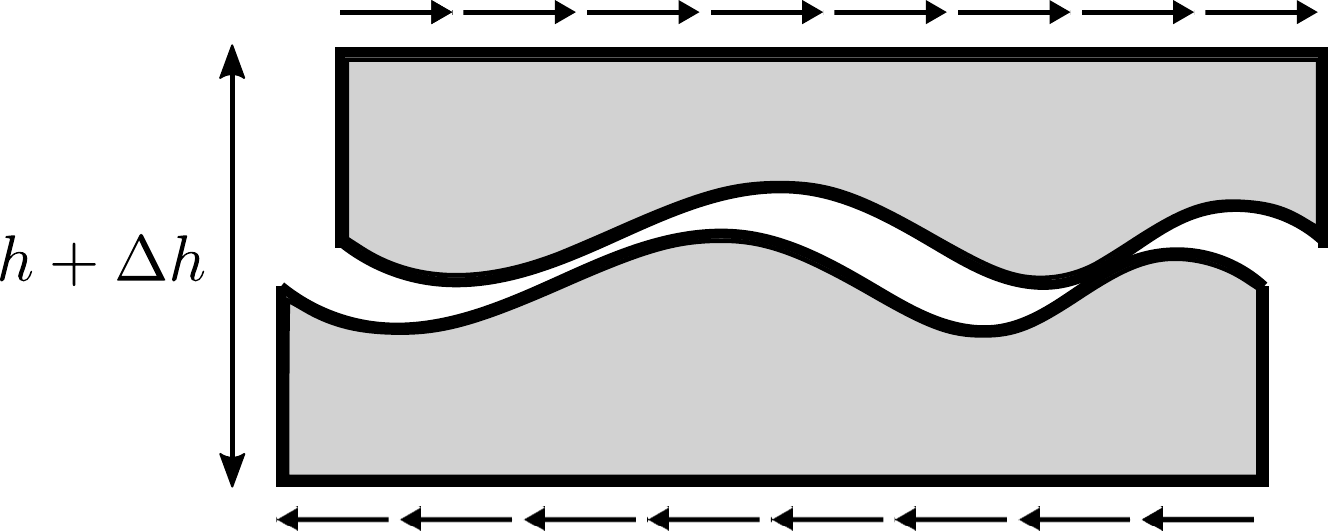}
    \caption{}
    \label{}
    \end{subfigure}
    \caption{Sketch showcasing the dilatancy effect on geomaterials, also known as Reynolds dilatancy. Bulk expansion takes place due to the lever motion that occurs between neighbouring grains as a result of interlocking.}
    \label{fig:Dilatancy}
\end{figure}

We begin our numerical experiments by conducting shear tests on a single element. The aim is to investigate the ability of the Drucker-Prager based formulation presented in capturing frictional behaviour and the dilatancy effect. The latter is the volume change observed in granular materials subjected to shear deformations, due to the interlocking between grains and interfaces (see Fig. \ref{fig:Dilatancy}). 

\begin{figure}[H]
    \centering
    \includegraphics[width=0.2\textwidth]{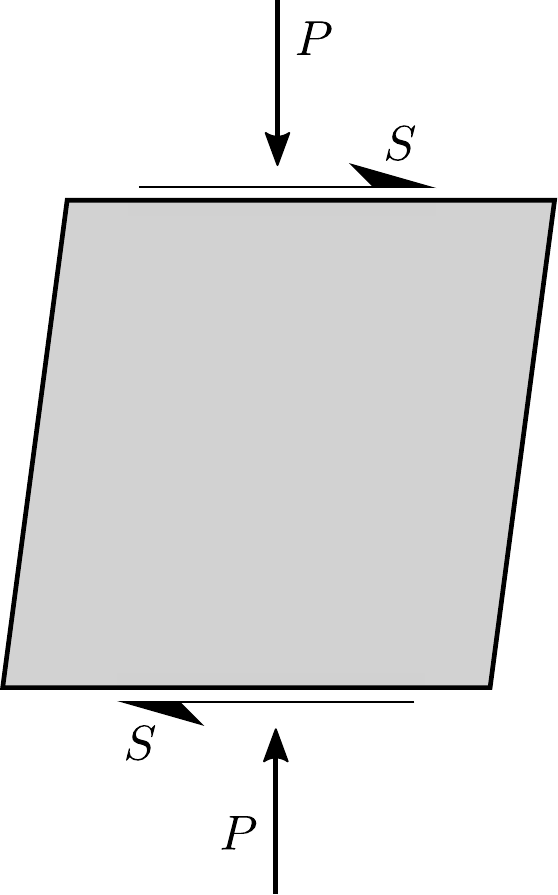}
    \caption{Configuration of a single element under pressure and shear
stress.}
    \label{fig:single-config}
\end{figure}

As shown in Fig. \ref{fig:single-config}, a single plane strain element is considered undergoing both shear and uniaxial pressure. Specifically, a vertical constant pressure is first applied, followed by shear displacement at the top and bottom edges. In this and all other case studies, the Neumann boundary condition $\nabla \phi \cdot \mathbf{n}=0$ is adopted for the phase field. The constitutive behaviour of the element is characterised by linear elasticity, with a Young's modulus of $E=25$ GPa and a Poisson’s ratio of $\nu=0.2$. The fracture behaviour is described by a material toughness of $G_c=0.15$ kJ/m$^2$ and a phase field length scale of $\ell=2$ mm. 

We aim at assessing the frictional behaviour of the model, for which it is convenient to formulate the relation between the shear strain $\varepsilon_{x y}$ and the shear stress $\sigma_{x y}$, as a function of the pressure and Drucker-Prager's $B$ parameter. For the fully damaged state ($\phi=1$), this relation reads
\begin{equation}
  (\sigma_f)_{xy}=\frac{\partial \psi_c(\boldsymbol\varepsilon)}{\partial \varepsilon_{xy}}=
\frac{K \mu}{9 B^2 K +  \mu} \left(\frac{I_1(\boldsymbol\varepsilon)}{\sqrt{J_2}(\boldsymbol\varepsilon)}+6 B \right) \varepsilon_{xy} 
\label{Eq:Exy-Sxy}
\end{equation}

\begin{figure}[H]
    \centering
    \begin{subfigure}[t]{0.45\textwidth}
    \includegraphics[width=\textwidth]{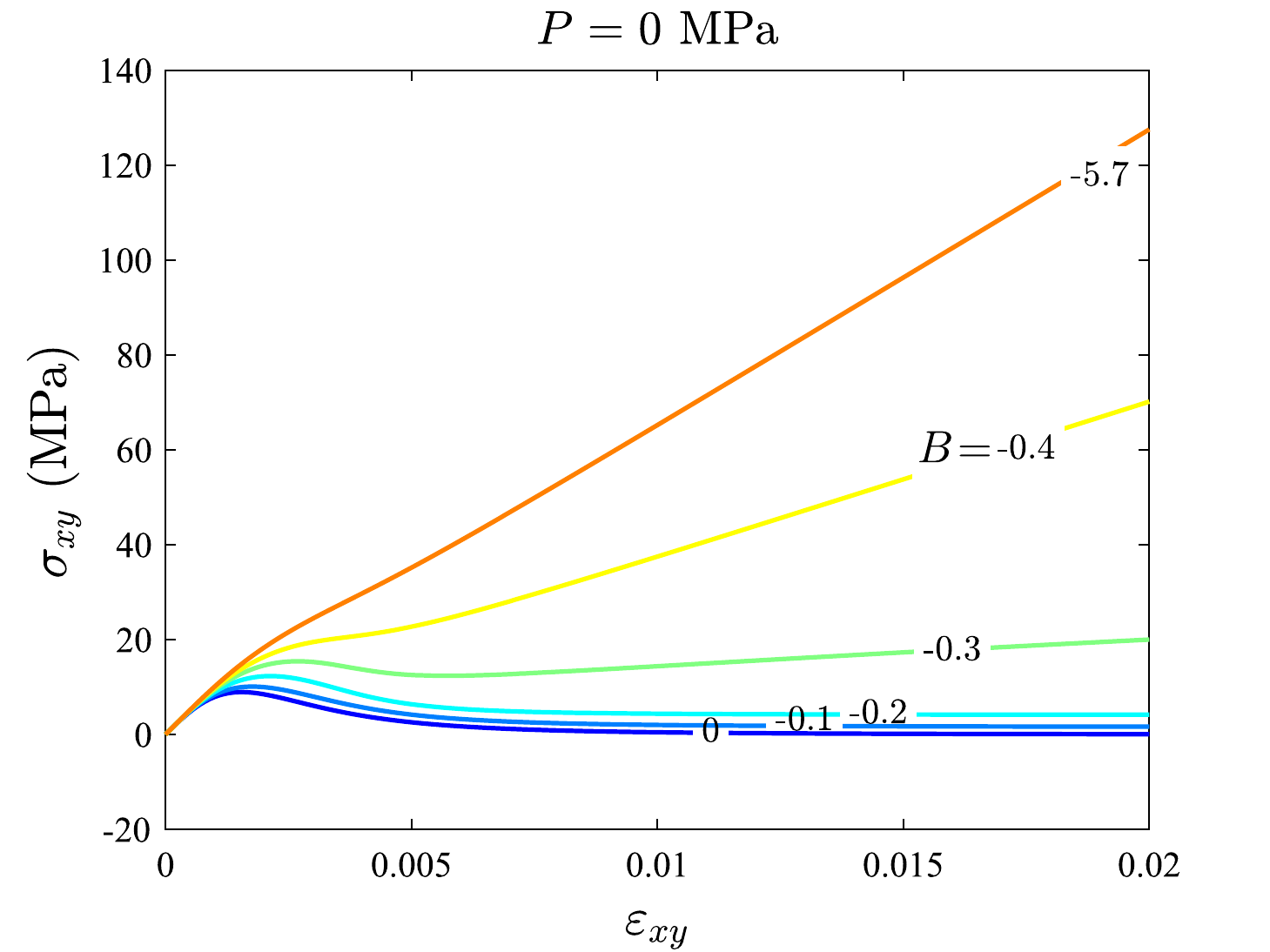}
    \caption{}
    \label{fig:Single-P-0-a}
    \end{subfigure} 
    \begin{subfigure}[t]{0.465\textwidth} 
    \includegraphics[width=\textwidth]{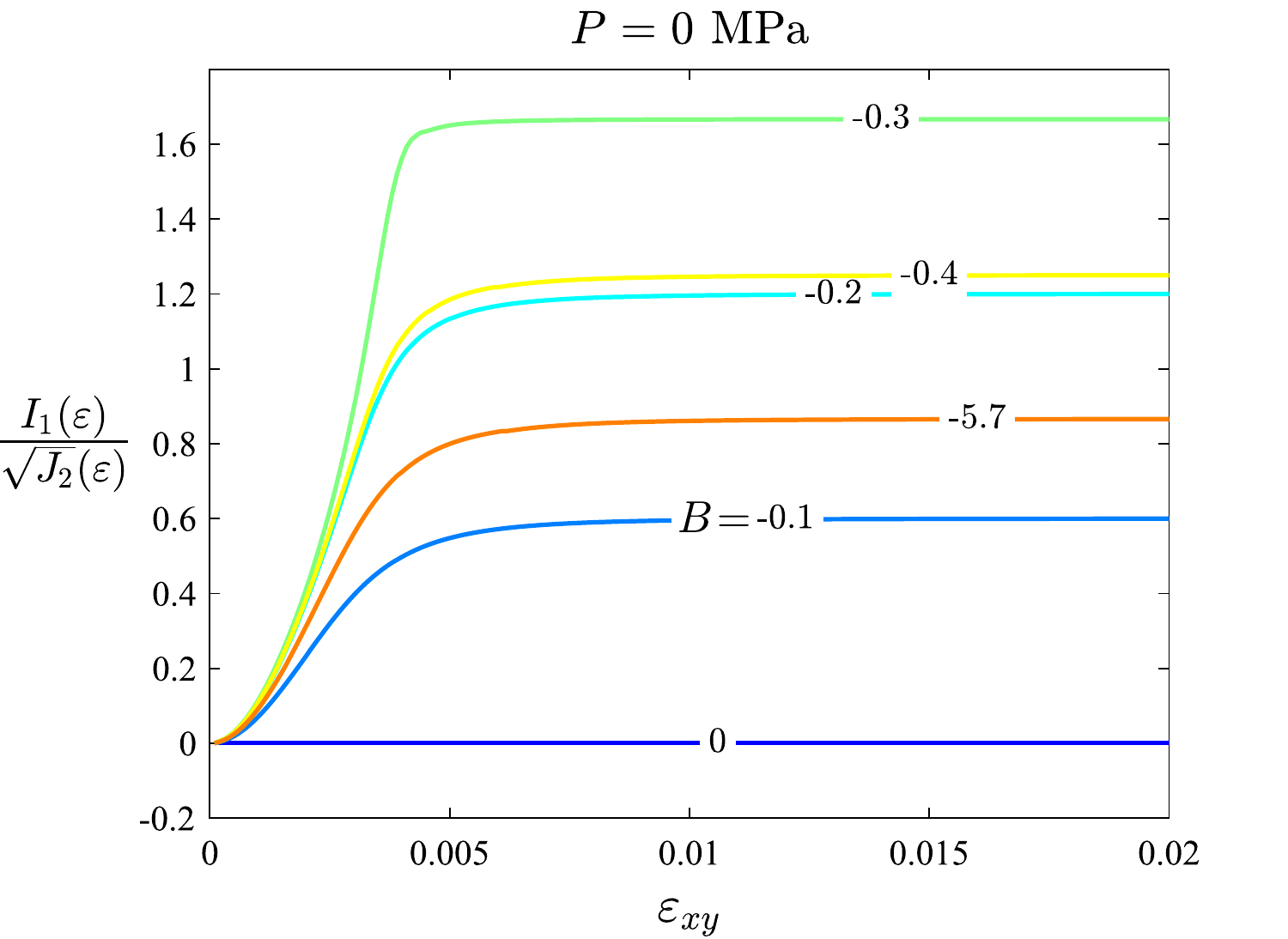}
    \caption{}
    \label{fig:Single-P-0-b}
    \end{subfigure}
    \caption{Single element under shear deformation. Results obtained without vertical pressure ($P=0$ MPa) for selected choices of $B$: (a) shear stress $\sigma_{x y}$ versus shear strain $\varepsilon_{x y}$, and (b) $I_1(\boldsymbol{\varepsilon})/ \sqrt{J_2}(\boldsymbol{\varepsilon})$ versus shear strain $\varepsilon_{x y}$.}
    \label{fig:Single-P-0}
\end{figure}

First, let us consider the case of no pressure ($P=0$). Fig. \ref{fig:Single-P-0-a} shows the shear stress versus shear strain curves obtained for different $B$ values. The role played by damage evolution can be readily observed, with calculations obtained for low absolute $B$ values exhibiting a peak in the shear stress response. For the fully cracked state ($\phi=1$), the shear stress drops to zero only if $B=0$. Hence, the expected influence of dilatancy on the stress-strain curve is attained for $B\neq 0$, and the effect increases with increasing its absolute magnitude ($|B|$). This load bearing capacity that is retained after reaching the fully cracked state due to dilatancy arises due to two contributions. One is the term $6B$ in Eq. (\ref{Eq:Exy-Sxy}). The second one is the term $I_1(\boldsymbol\varepsilon)/\sqrt{J_2}(\boldsymbol\varepsilon)$ - as shown in Fig. \ref{fig:Single-P-0-b}, it attains a positive constant value for $\phi=1$ and $B\neq0$. However, the relation between $B$ and $I_1(\boldsymbol\varepsilon)/\sqrt{J_2}(\boldsymbol\varepsilon)$ is non-linear.

\begin{figure}[H]
    \centering
    \begin{subfigure}[t]{0.48\textwidth}
    \includegraphics[width=\textwidth]{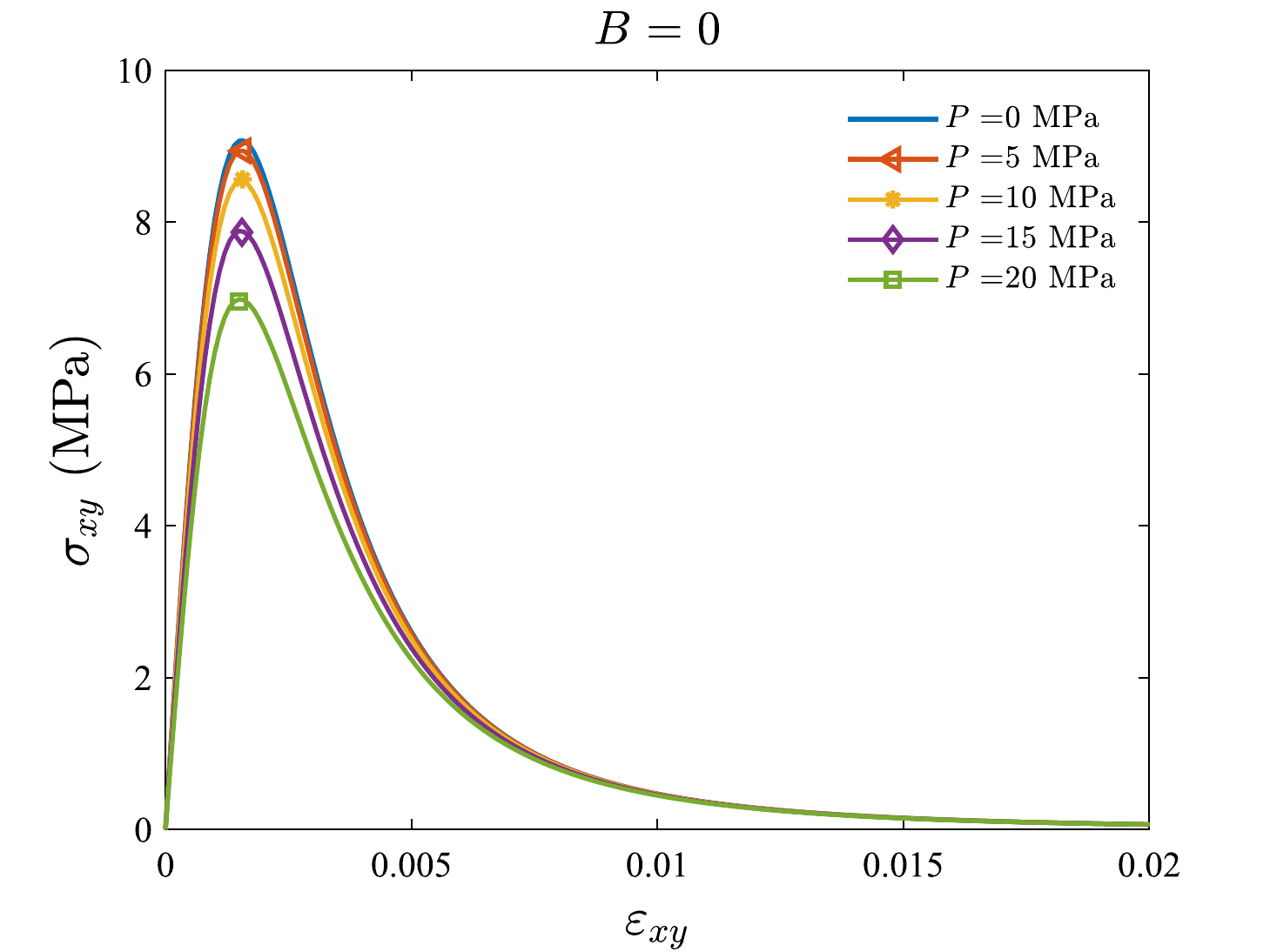}
    \caption{}
    \label{fig:Sxy-Exy-B-a}
    \end{subfigure} 
    \begin{subfigure}[t]{0.48\textwidth} 
    \includegraphics[width=\textwidth]{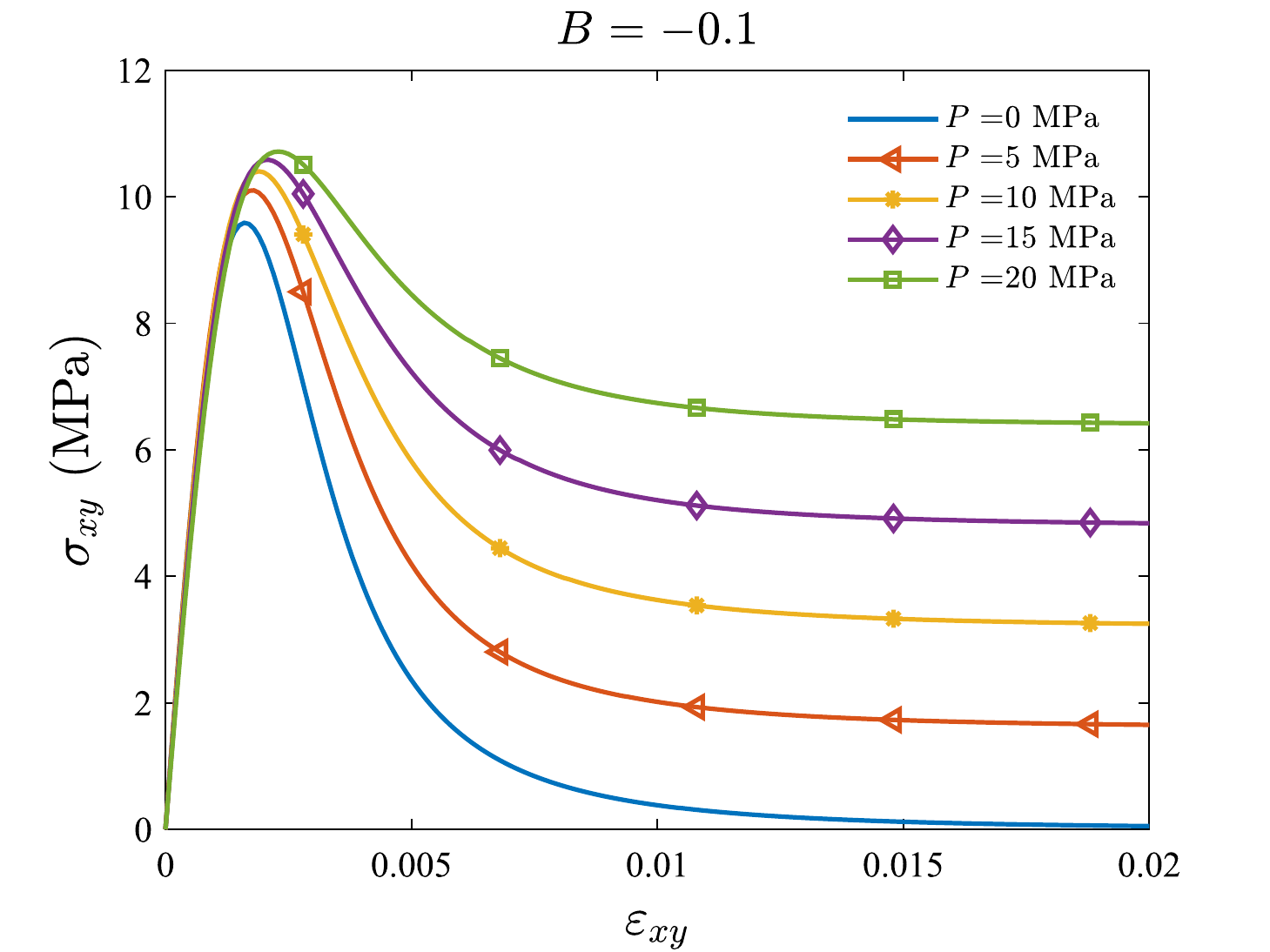}
    \caption{}
    \label{fig:Sxy-Exy-B-b}
    \end{subfigure}
    \begin{subfigure}[t]{0.48\textwidth} 
    \includegraphics[width=\textwidth]{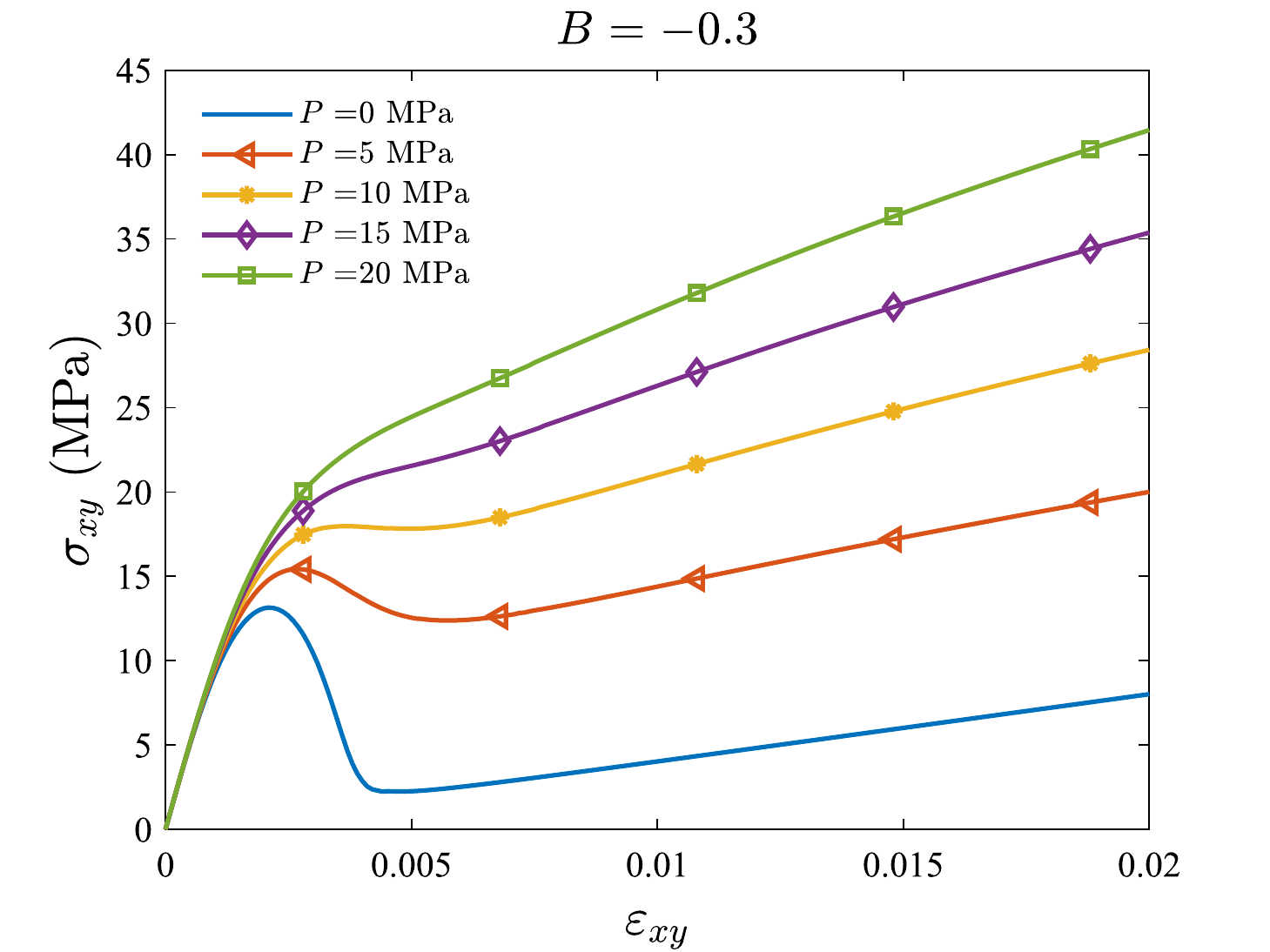}
    \caption{}
    \label{fig:Sxy-Exy-B-c}
    \end{subfigure}
    \begin{subfigure}[t]{0.48\textwidth} 
    \includegraphics[width=\textwidth]{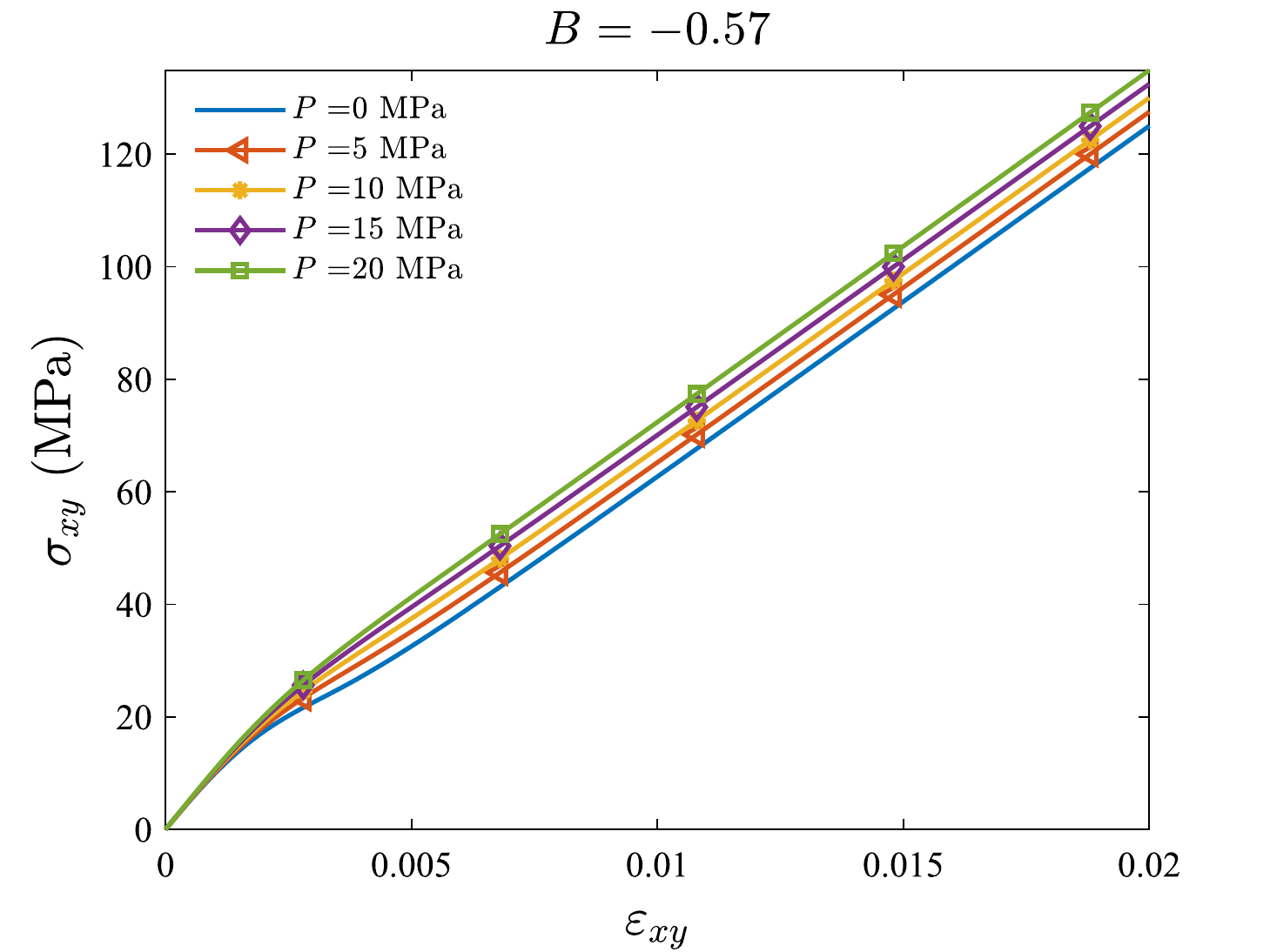}
    \caption{}
    \label{fig:Sxy-Exy-B-d}
    \end{subfigure}
    \caption{Single element under shear deformation. Shear stress versus shear strain predictions as a function of $P$ for selected values of the $B$ parameter: (a) $B=0$, (b) $B=-0.1$, (c) $B=-0.3$, and (c) $B=-0.57$.}
    \label{fig:Sxy-Exy-B}
\end{figure}

Next, the influence of vertical pressure is examined. The results obtained for selected values of $P$ and $B$ are shown in Fig. \ref{fig:Sxy-Exy-B}. For the case of $B=0$ (Fig. \ref{fig:Sxy-Exy-B-a}), the shear stress shows a negligible sensitivity to the vertical pressure and no frictional effect ($\sigma_{xy}$ drops to zero as $\phi \to 1$). The peak stress value shows some sensitivity to $P$ due to the interplay between damage and the applied pressure. The results seen for $B=0$ contrast with those obtained for non-zero $B$ values (Figs. \ref{fig:Sxy-Exy-B}b-d). For $B \neq 0 $, friction plays a noticeable role with the shear stress increasing with $P$. Also, the slope of the shear stress-strain curve increases with the absolute value of $B$. % Why? we are missing a bit of interpretation here

\begin{figure}[H]
    \centering
    \begin{subfigure}[t]{0.48\textwidth}
    \includegraphics[width=\textwidth]{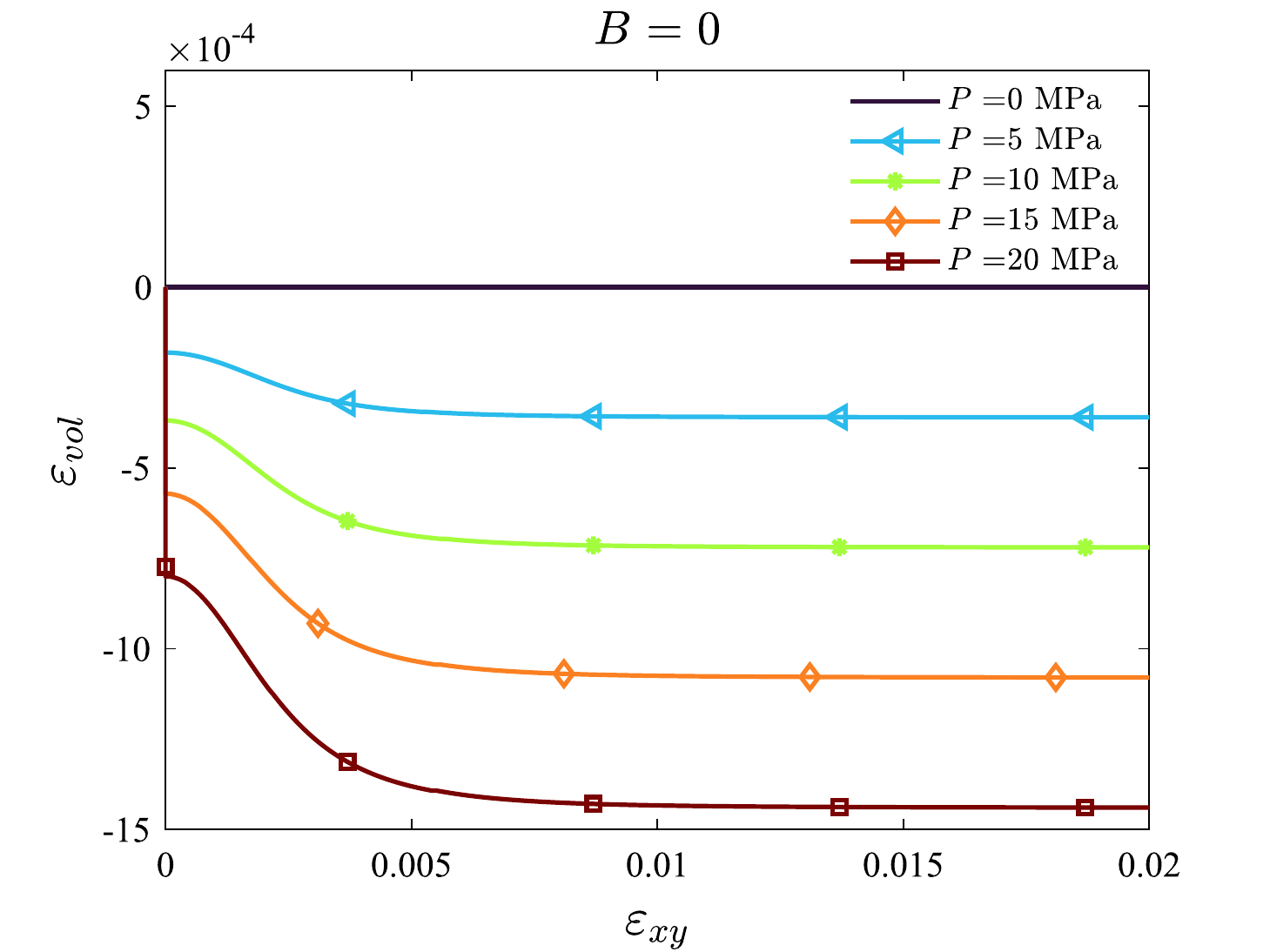}
    \caption{}
    \label{fig:single-Evol-Exy-a}
    \end{subfigure} 
    \begin{subfigure}[t]{0.48\textwidth} 
    \includegraphics[width=\textwidth]{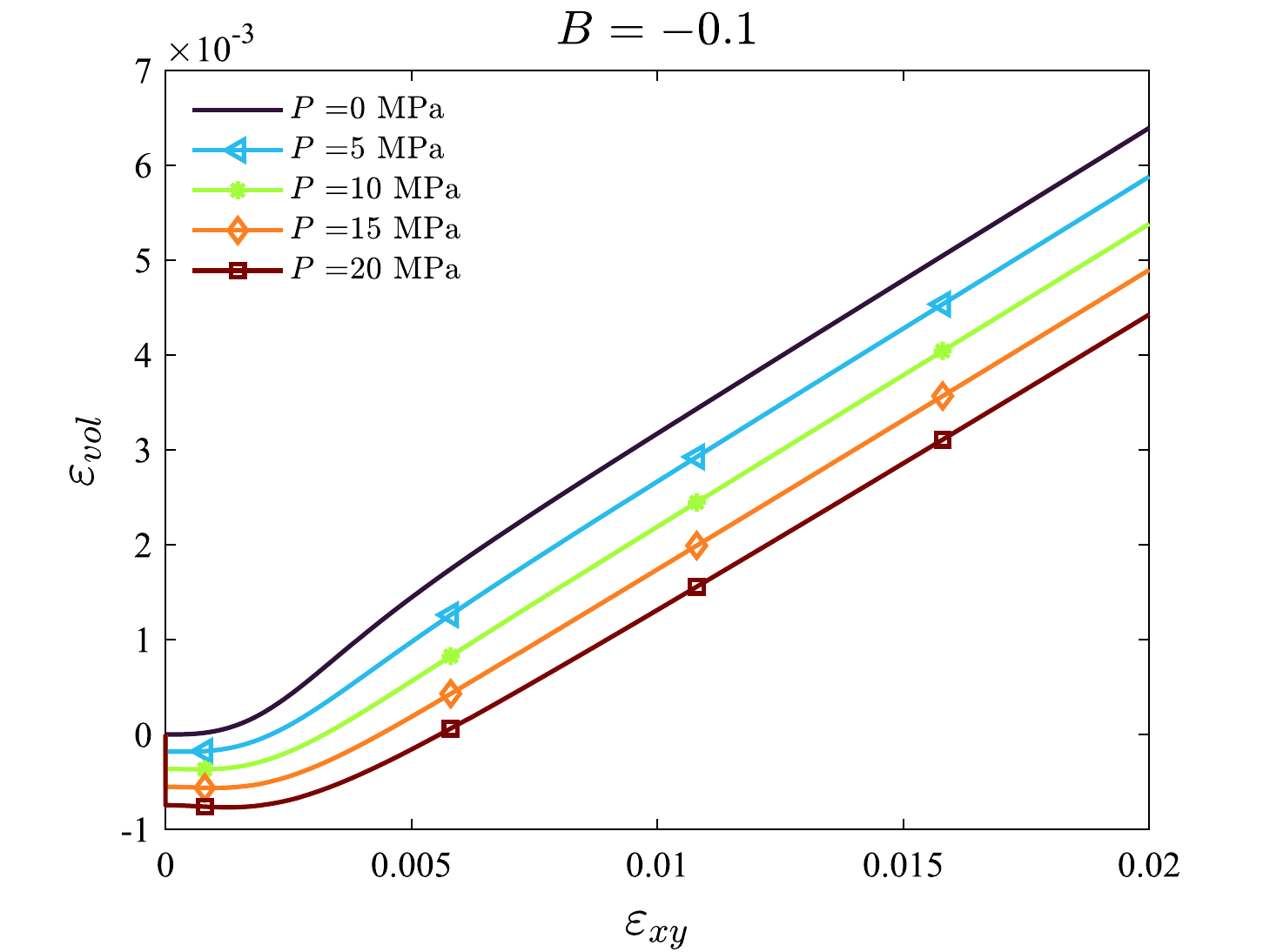}
    \caption{}
    \label{fig:single-Evol-Exy-b}
    \end{subfigure}
    \begin{subfigure}[t]{0.48\textwidth} 
    \includegraphics[width=\textwidth]{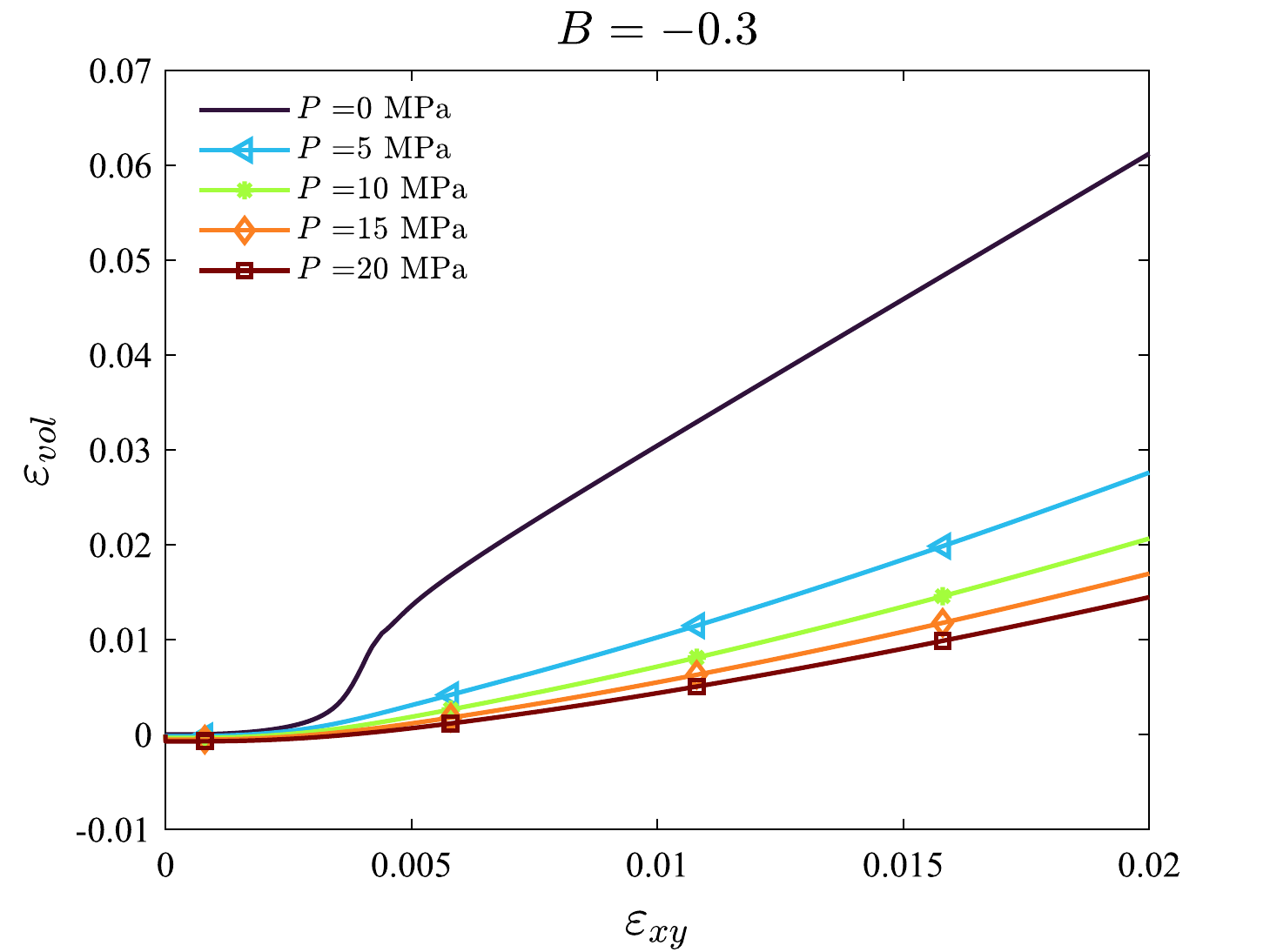}
    \caption{}
    \label{fig:single-Evol-Exy-c}
    \end{subfigure}
    \begin{subfigure}[t]{0.48\textwidth} 
    \includegraphics[width=\textwidth]{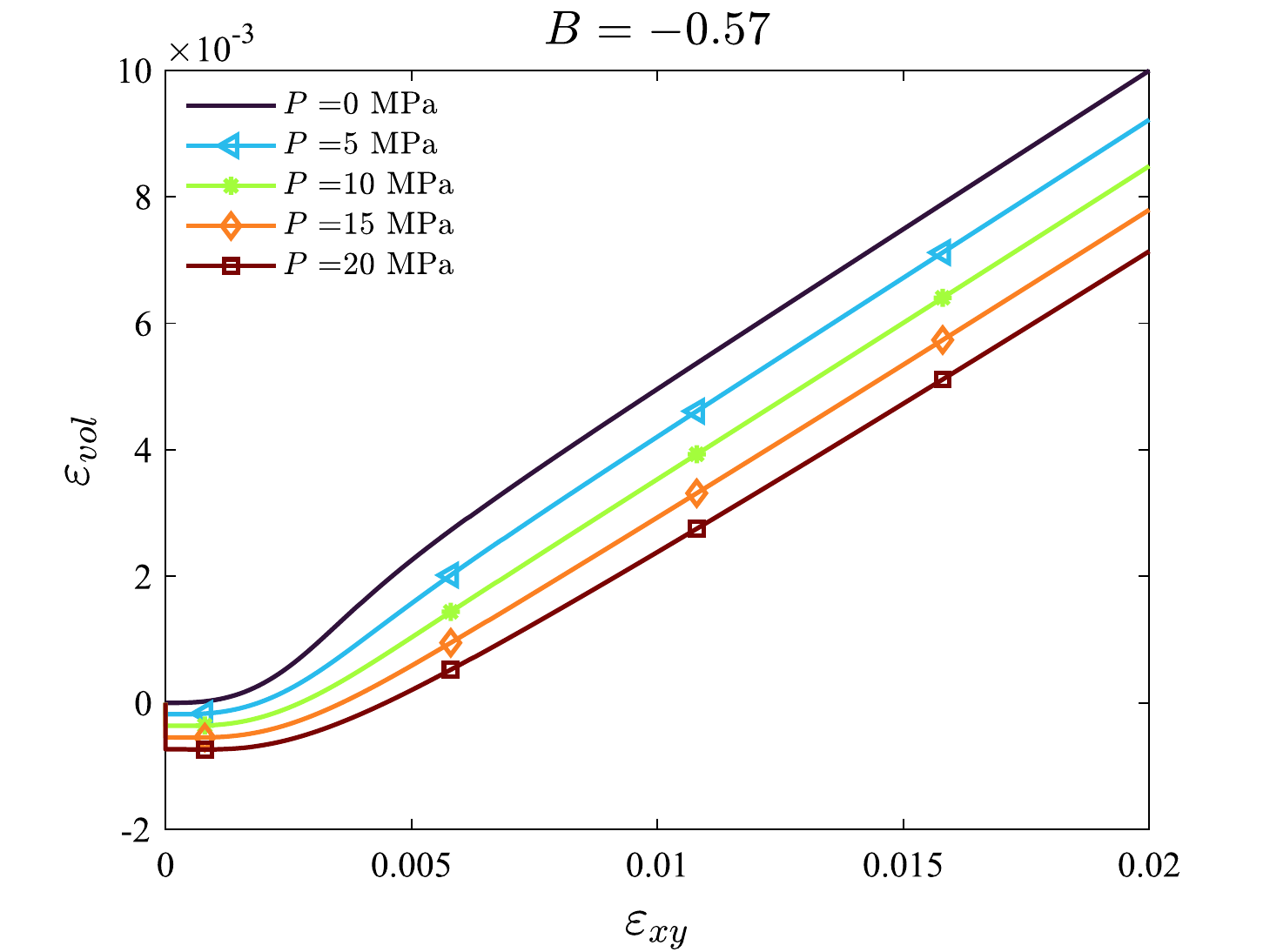}
    \caption{}
    \label{fig:single-Evol-Exy-d}
    \end{subfigure}
    \caption{Single element under shear deformation. Volumetric strain versus shear strain predictions as a function of $P$ for selected values of the $B$ parameter: (a) $B=0$, (b) $B=-0.1$, (c) $B=-0.3$, and (c) $B=-0.57$.}
    \label{fig:single-Evol-Exy}
\end{figure}

The ability of the Drucker-Prager based split model to capture the dilatancy effect is further explored by plotting the predictions of volumetric strain $\varepsilon_{vol}=\varepsilon_{xx}+\varepsilon_{yy}+\varepsilon_{zz}$ for selected values of the parameter $B$ and the applied pressure $P$. As shown in Fig. \ref{fig:single-Evol-Exy}, the volumetric strain $\varepsilon_{vol}$ increases with the shear strain $\varepsilon_{x y}$ in all cases except for that of $B=0$. The effect of dilatancy is clear in all $B \neq 0$ calculations (Fig. \ref{fig:single-Evol-Exy}b-d). In addition, the results show that higher pressures lead to reductions in volume as a result of material damage.

\subsection{\textit{Virtual} Direct Shear Tests (DST)}
\label{Sec:RDST}

\begin{figure}[H]
    \centering
    \includegraphics[width=0.5\textwidth]{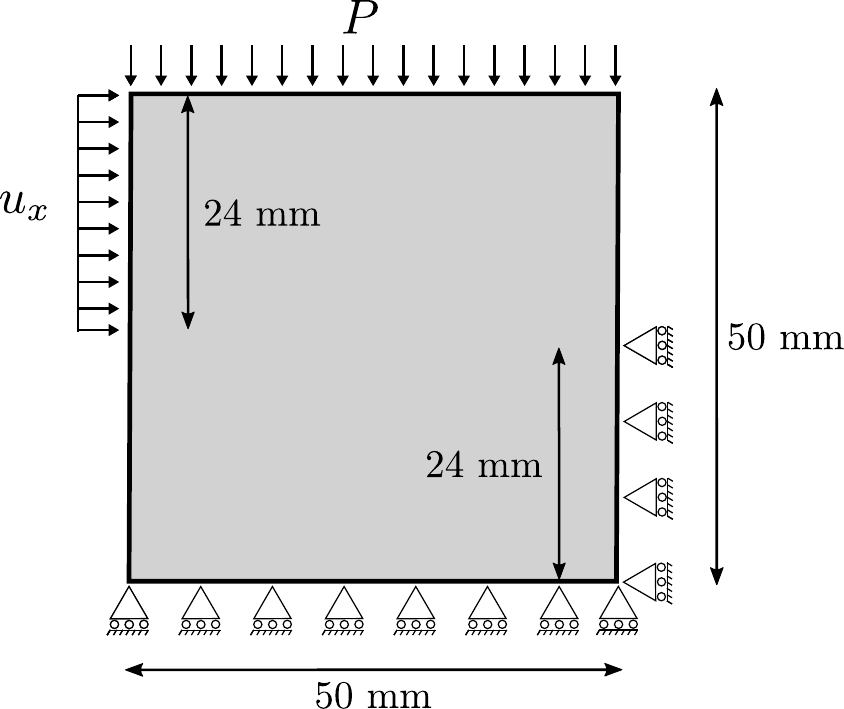}
    \caption{Direct shear test (DST) model. Geometry and boundary conditions.}
    \label{fig:DST-config}
\end{figure}

Next, the Direct Shear Test (DST) is simulated to evaluate the model behaviour in an experimental configuration that is widely used for finding the frictional parameters of soil and rock materials, such as cohesion and friction angle. The geometry and boundary conditions of the model are shown in Fig. \ref{fig:DST-config}. A vertical pressure $P$ is applied at the top edge, followed by a horizontal displacement $u_x$ over a 24 mm long region of the left edge. We consider three scenarios to assess the role of the vertical pressure: $P=20$ MPa, $P=10$ MPa and no pressure ($P=0$). The elastic properties are taken as $E=25$ GPa and $\nu=0.2$, while the fracture parameters are given by $G_c=0.15$ kJ/m$^2$ and $\ell=0.2$ mm. The model is discretised with approximately 80,000 4-node plane strain quadrilateral elements with reduced integration. The mesh is refined along the expected crack propagation region, such that the characteristic element size is at least half of the phase field length scale $\ell$. 

\begin{figure}[H]
    \centering
    \includegraphics[width=0.7\textwidth]{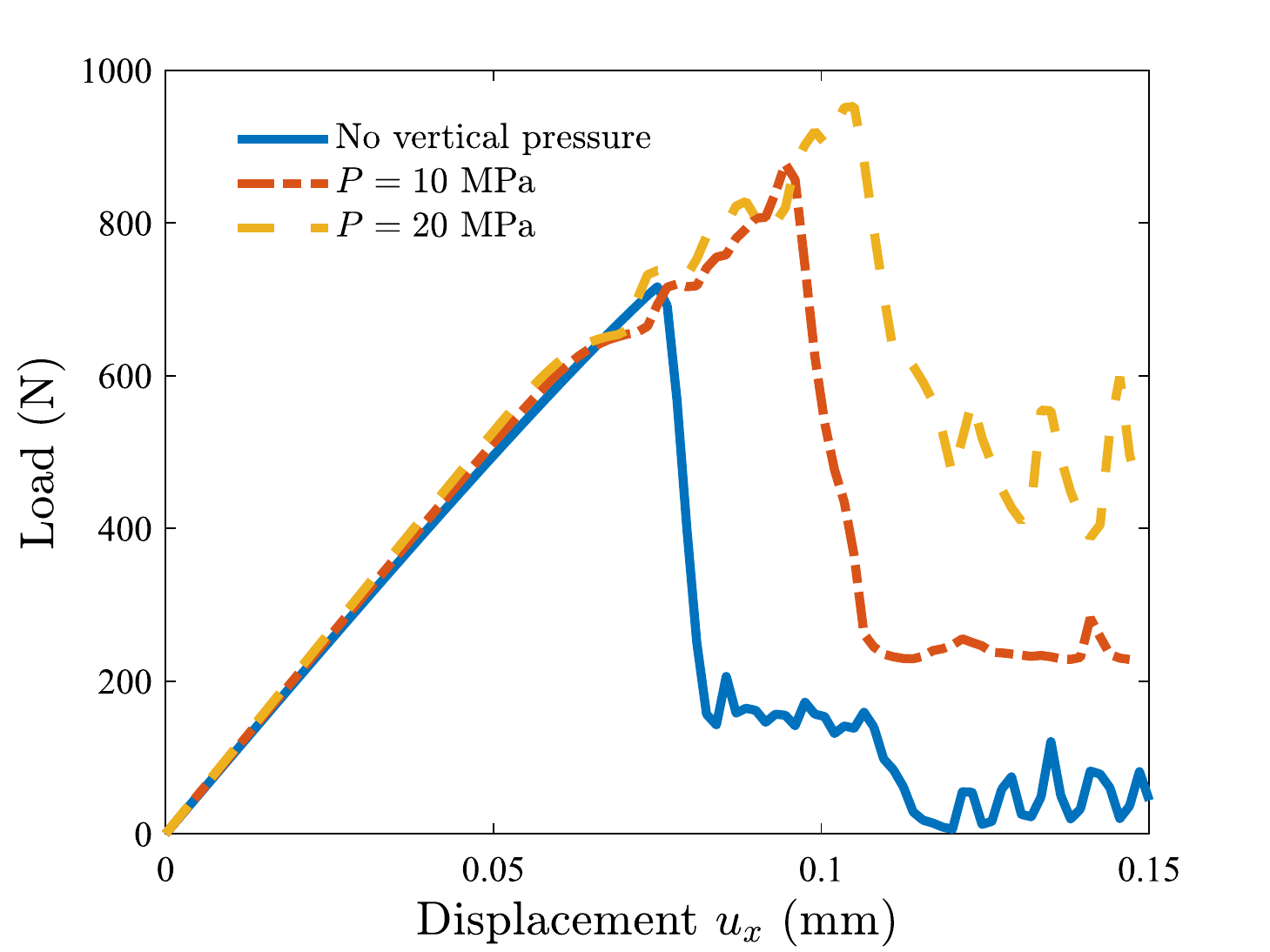}
    \caption{Direct shear test (DST). Shear load versus applied displacement results as a function of the applied pressure $P$.}
    \label{fig:DST-LD}
\end{figure}

The results obtained are shown in Fig. \ref{fig:DST-LD}, in terms of the shear force versus the applied displacement $u_x$, and as a function of the applied pressure $P$. The case of no pressure shows a complete drop of the load carrying capacity as a result of damage, in agreement with experimental DST observations on geomaterials. However, a residual load is retained when a vertical pressure is applied, and this increases with the magnitude of $P$. Also, in all cases some oscillations can be seen in the force versus displacement response, which can be attributed to the effect of grain interlocking.  

\begin{figure}[H]
    \centering
    \begin{subfigure}[t]{0.25\textwidth} 
    \includegraphics[width=\textwidth]{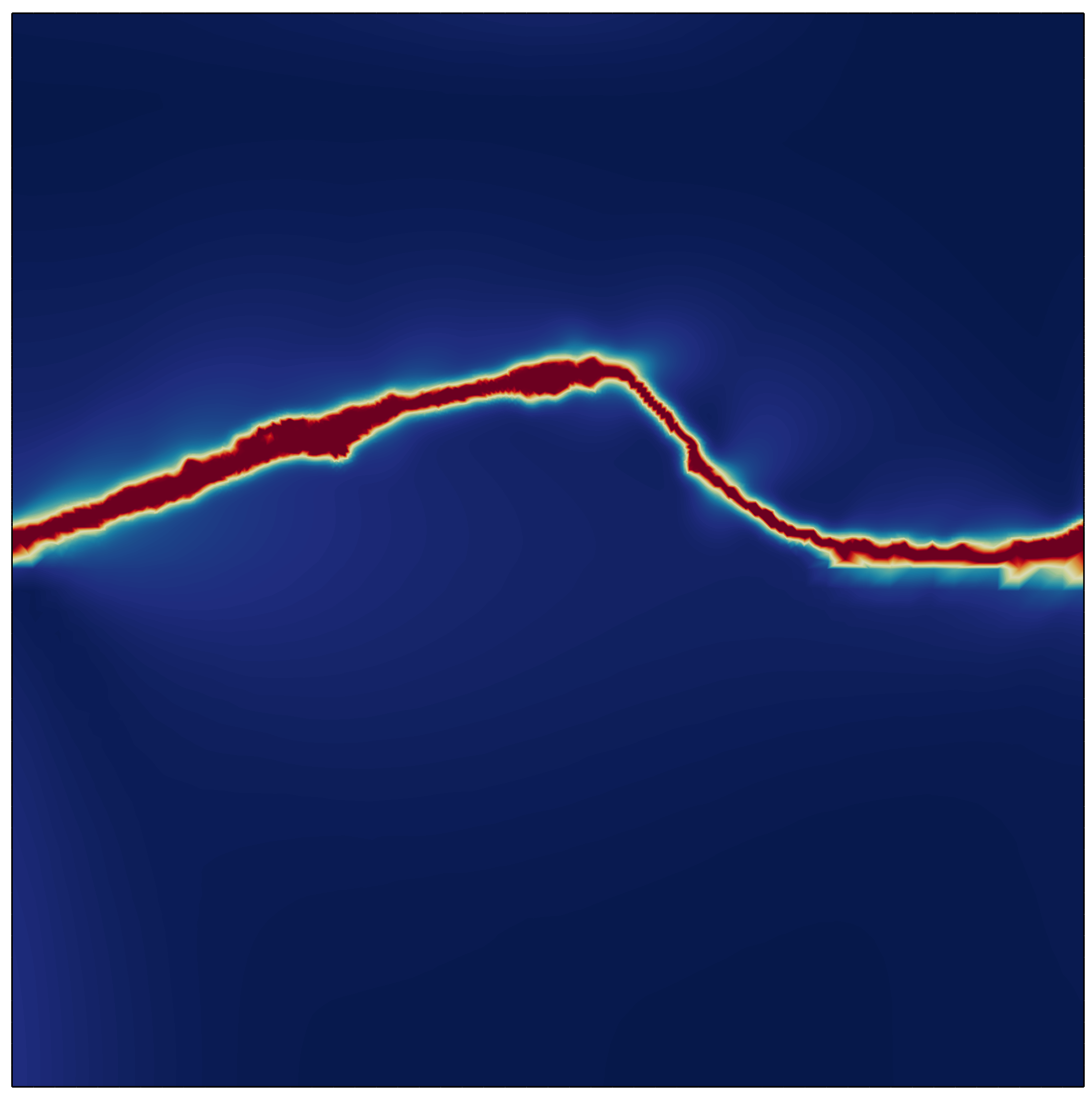} 
    \caption{$P=0$ MPa}
    \label{}
    \end{subfigure} \hspace{.5 cm}
    \begin{subfigure}[t]{0.25\textwidth} 
    \includegraphics[width=\textwidth]{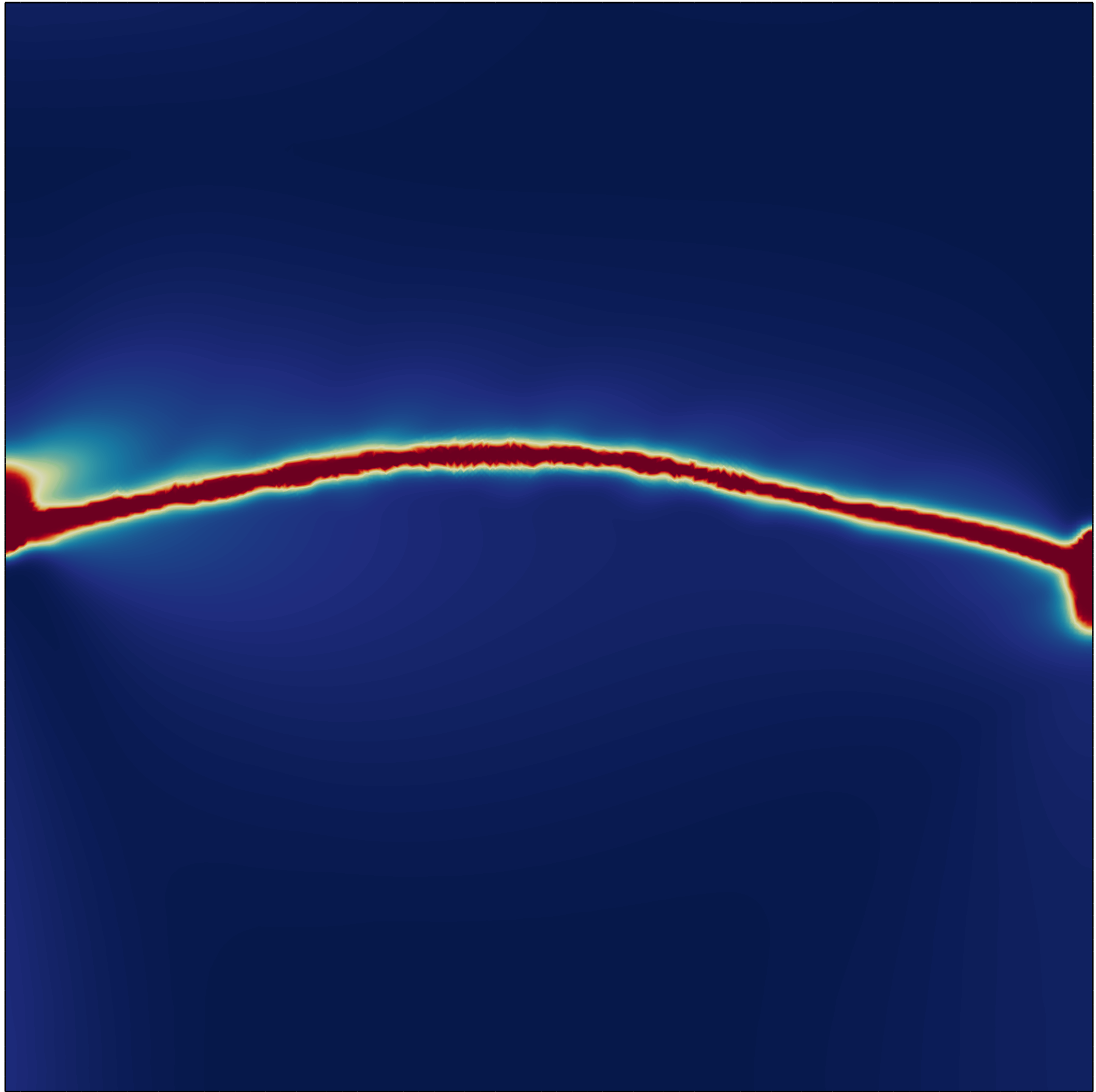}
    \caption{$P=10$ MPa}
    \label{}
    \end{subfigure}\hspace{.5 cm}
    \begin{subfigure}[t]{0.25\textwidth} 
    \includegraphics[width=\textwidth]{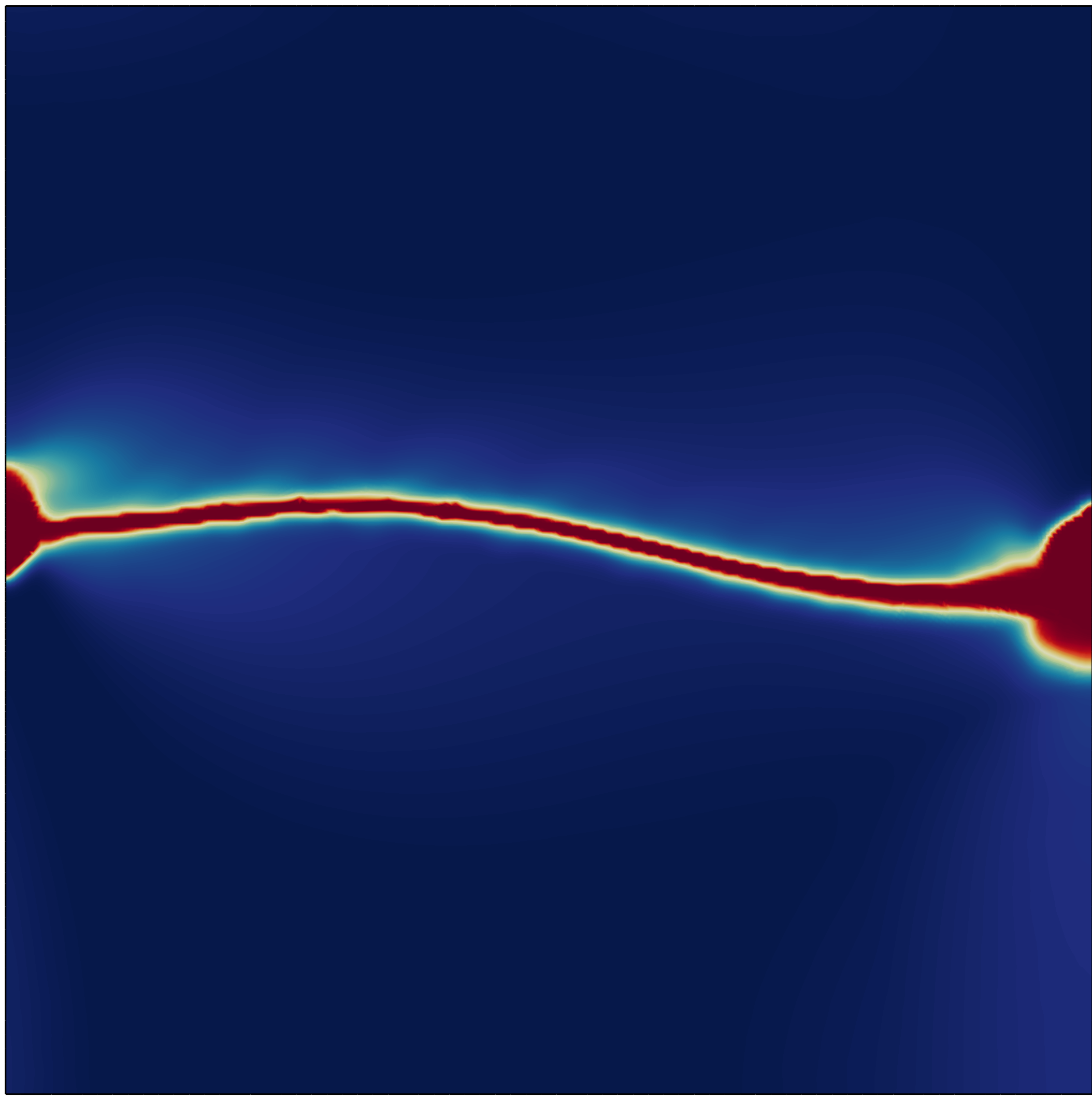}
    \caption{$P=20$ MPa}
    \label{}
    \end{subfigure}
    \begin{subfigure}[t]{0.1\textwidth}
    \includegraphics[width=\textwidth]{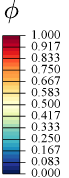}
    \end{subfigure}
    \caption{Direct shear test (DST). Predicted cracking patterns, as shown through contours of the phase field $\phi$ for selected values of the applied pressure: (a) $P=0$, (b) $P=10$ MPa, and (c) $P=20$ MPa.}
    \label{fig:phi-DST}
\end{figure}

Finally, the predicted crack trajectories are shown in Fig. \ref{fig:phi-DST}, as a function of $P$, by plotting contours of the phase field order parameter $\phi$. The results reveal an influence of the applied pressure on the cracking pattern. The lower the vertical pressure the more tortuous the crack path. Also, increasing the applied pressure leads to an accumulation of damage at the edges of the loading region, which are then connected through a crack that propagates across the sample.

\subsection{Uniaxial and triaxial compression testing of concrete}
\label{Sec:Rcompressionconcrete}

\begin{figure}[H]
    \centering
    \includegraphics[width=0.4\textwidth]{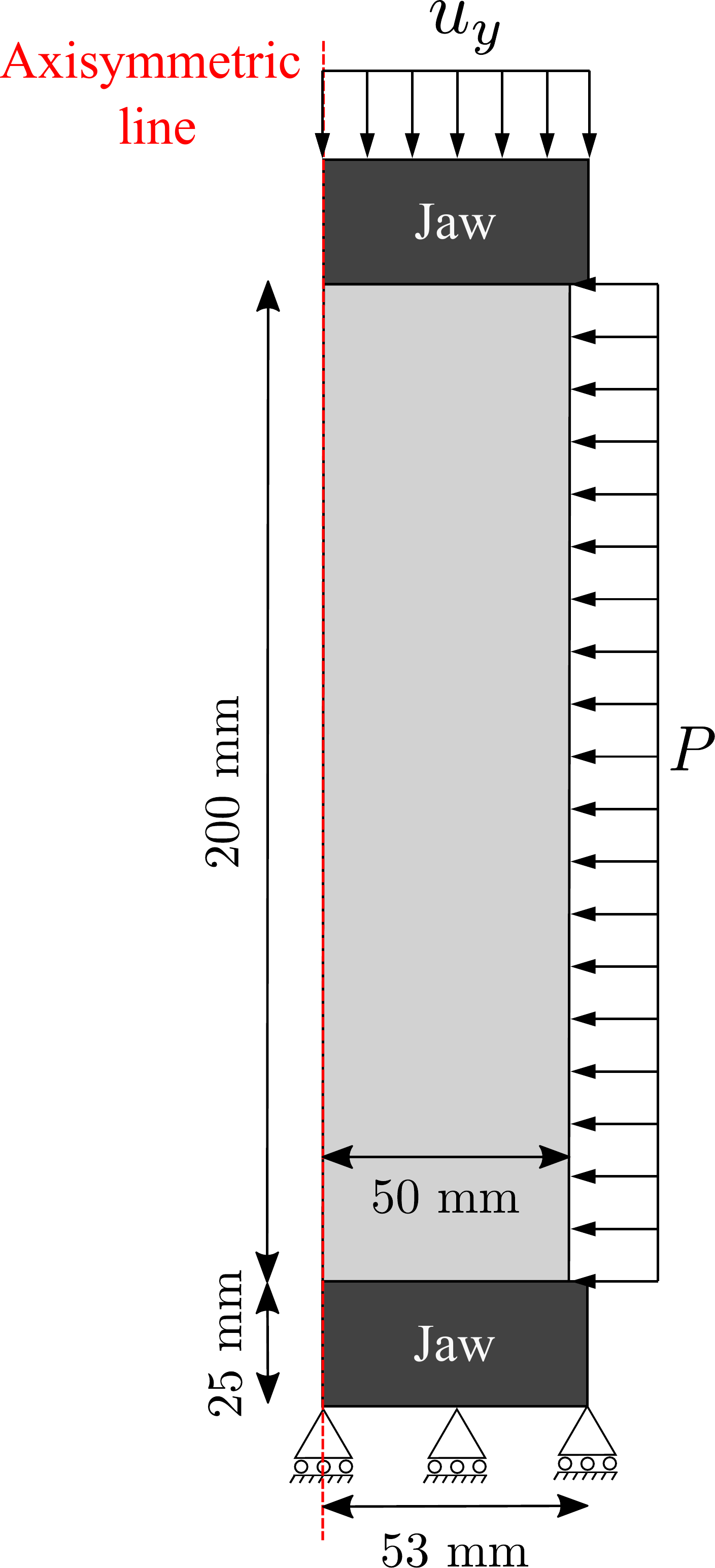}
    \caption{Compressive failure of concrete. Model geometry, dimensions and boundary conditions.}
    \label{fig:CT-config}
\end{figure}

The third case study involves the failure of concrete samples undergoing uniaxial and triaxial compression. The aim is to investigate the abilities of the Drucker-Prager formulation presented to capture the effect of confinement. Mimicking the commonly used experimental setup, a cylindrical specimen is subjected to a compressive displacement at the top, while its surface is subjected to a confinement pressure. In the numerical model, we take advantage of axial symmetry and simulate a 2D section of the sample. The dimensions and loading configuration of the model are given in Fig. \ref{fig:CT-config}. To reproduce with fidelity the experimental conditions, we choose to simulate the contact between the jaws and the concrete sample. The jaws are assumed to be made of steel, with elastic properties $E=210$ GPa and $\nu=0.3$. The contact between the jaws and the disk is defined as a surface to surface contact with a finite sliding formulation. The tangentional contact behaviour is assumed to be frictionless while the normal behaviour is based on a hard contact scheme, where the contact constraint is enforced with a Lagrange multiplier representing the contact pressure in a mixed formulation. The material properties of concrete are taken to be $E=25$ GPa, $\nu=0.2$, $\ell=0.4$ mm, $G_c=0.15$ kJ/m$^2$, and $B=-0.12$. Linear quadrilateral axisymmetric elements are used to discretise the model. In particular, approximately 35,000 elements are used to discretise the concrete sample while 1,500 elements are employed in each of the jaws. The characteristic element size in the areas of interest is below 0.2 mm, half of the phase field length scale. The ratio between the applied pressure and the prescribed displacement equals $P/u_y=10$ MPa/mm. 

\begin{figure}[H]
    \centering
    \includegraphics[width=0.7\textwidth]{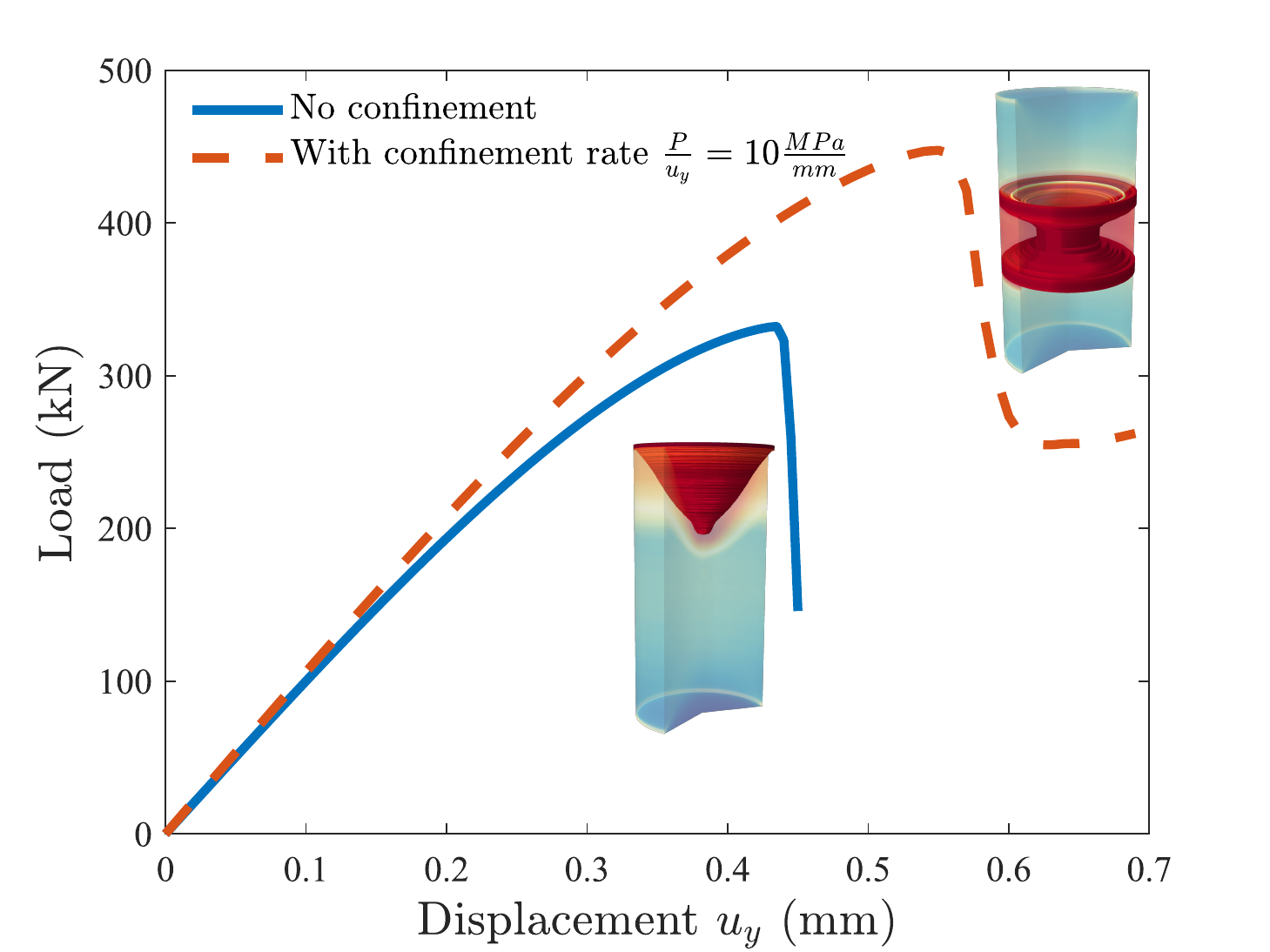}
    \caption{Compressive failure of concrete. Predicted load versus displacement curves for a sample without confinement pressure and one with a confinement pressure-prescribed displacement ratio of $P/u_y=10$ MPa/mm.}
    \label{fig:LLD-CT}
\end{figure}

The force versus displacement responses predicted with and without a confinement pressure are shown in Fig. \ref{fig:LLD-CT}. It can be seen that, in agreement with expectations, the application of a confinement pressure increases the magnitude of the critical load. The ultimate strength of the sample with confinement is found to be almost $40\%$ higher than the unconfined one. Also, a more brittle behaviour is observed in the unconfined sample, with a sharper drop in the load carrying capacity at the moment of failure. 

\begin{figure}[H]
    \centering
    \begin{subfigure}[t]{0.19\textwidth} 
    \includegraphics[width=\textwidth]{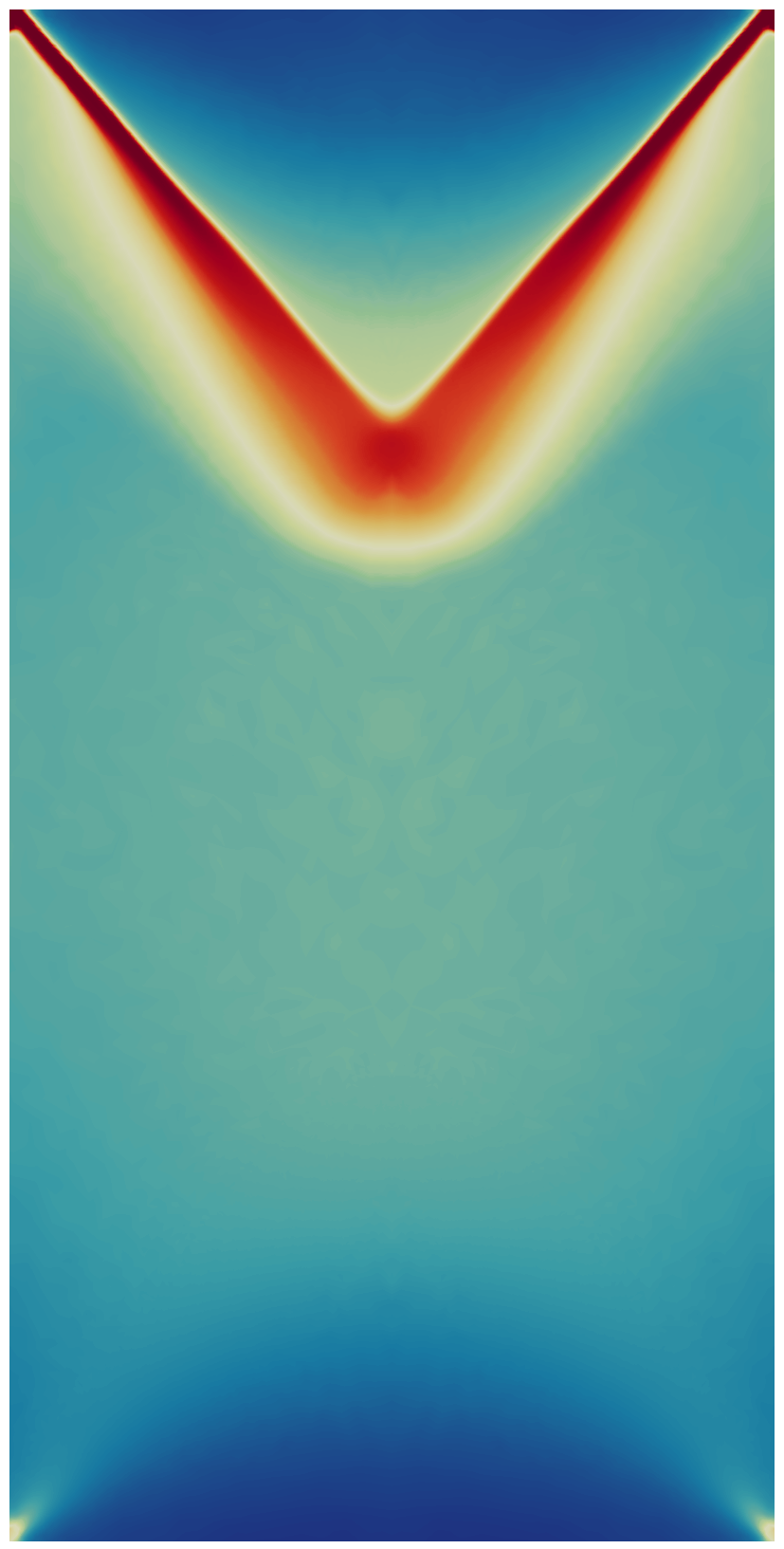} 
    \caption{}
    \label{}
    \end{subfigure} \hspace{1 cm}
    \begin{subfigure}[t]{0.2\textwidth} 
    \includegraphics[width=\textwidth]{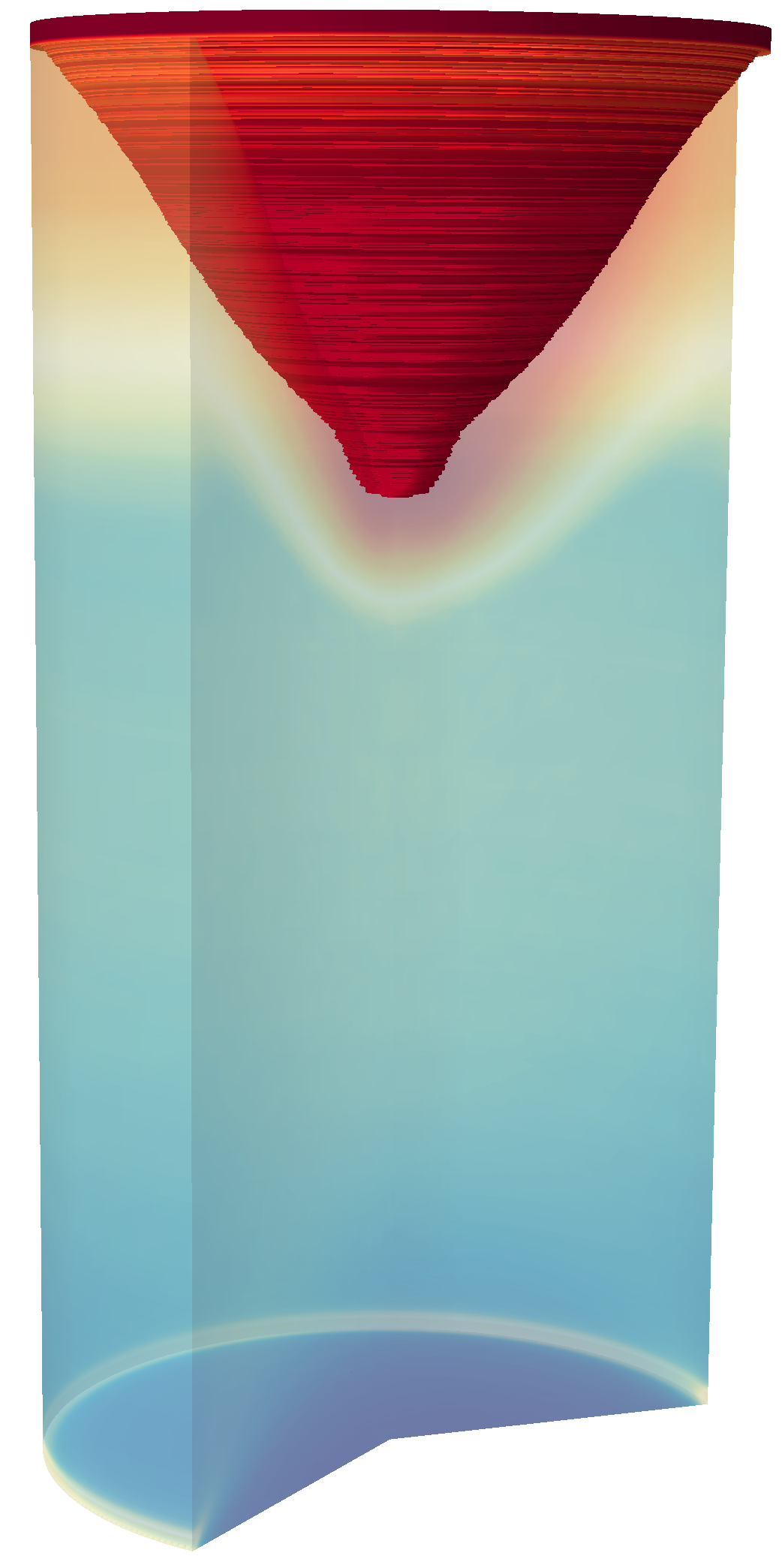}
    \caption{}
    \label{}
    \end{subfigure}\hspace{1 cm}
    \begin{subfigure}[t]{0.1\textwidth} 
    \includegraphics[width=\textwidth]{legend.pdf}
    \end{subfigure}
    \caption{Compressive failure of concrete. Cracking patterns for the unconfined sample, as described by the phase field $\phi$ contours: (a) axisymmetric 2D results, and (b) 3D visualisation.}
    \label{fig:CT-unconfined}
\end{figure}

\begin{figure}[H]
    \centering
    \begin{subfigure}[t]{0.19\textwidth} 
    \includegraphics[width=\textwidth]{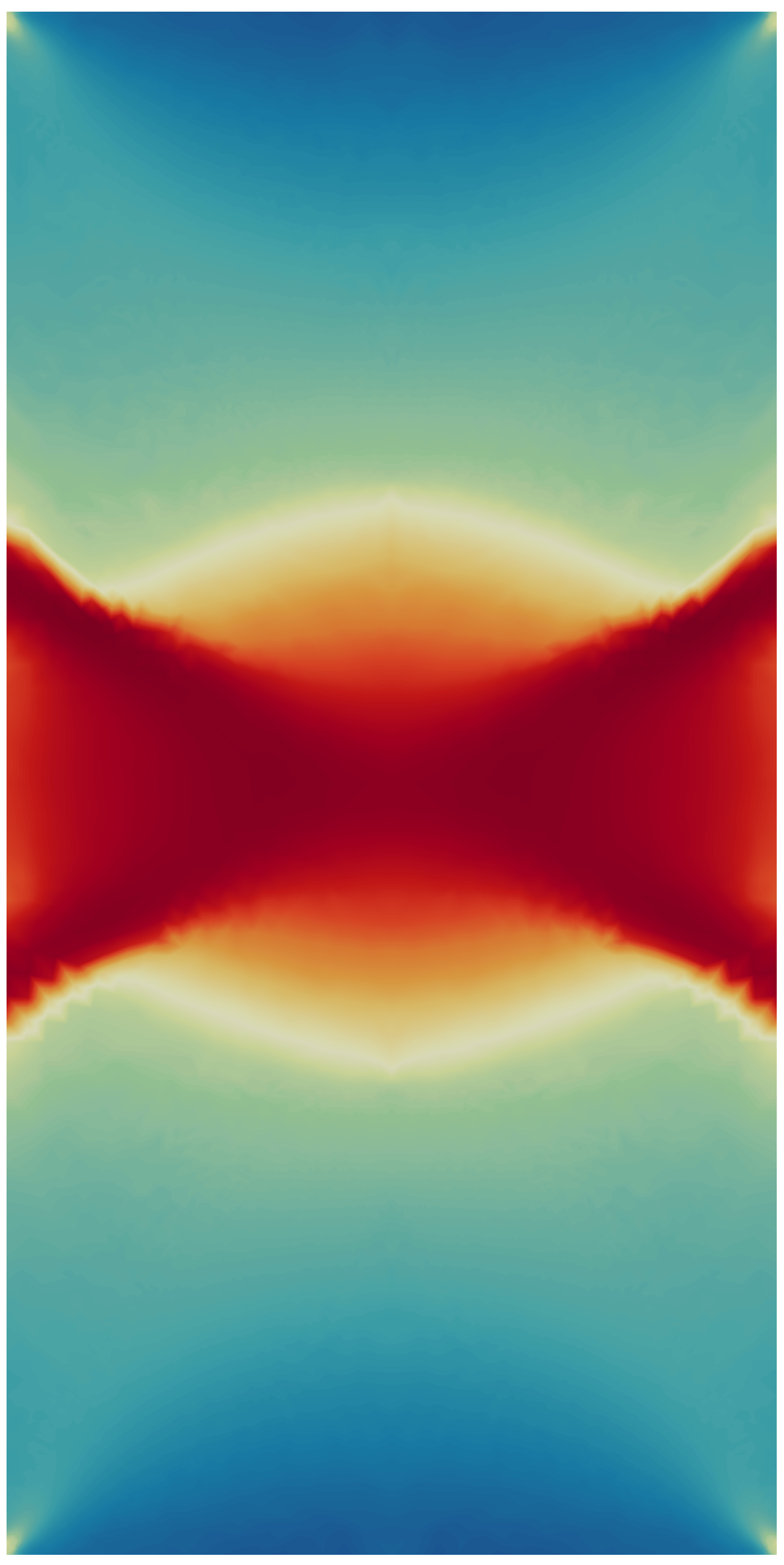} 
    \caption{}
    \label{}
    \end{subfigure} \hspace{1 cm}
    \begin{subfigure}[t]{0.2\textwidth} 
    \includegraphics[width=\textwidth]{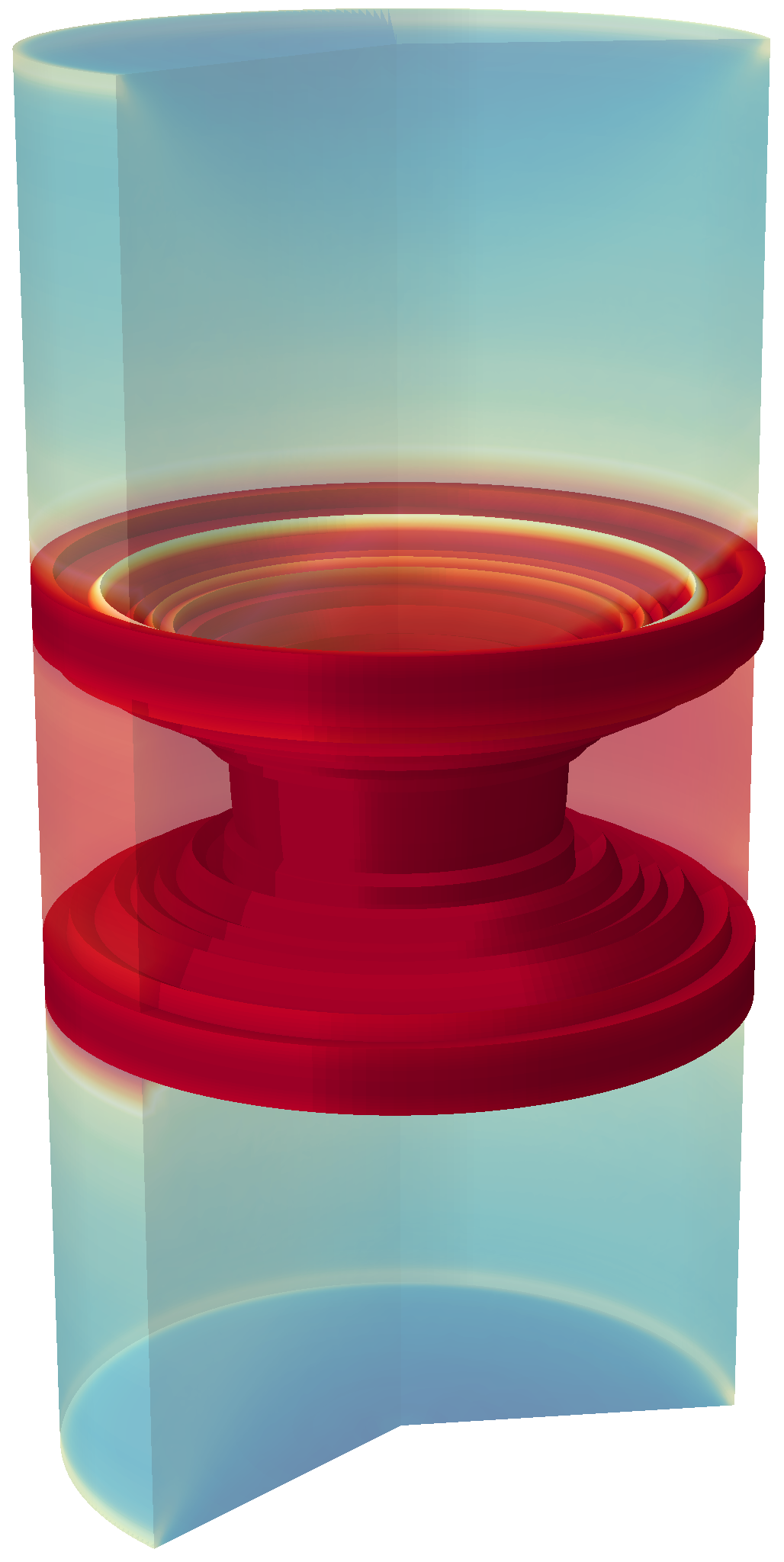}
    \caption{}
    \label{}
    \end{subfigure}\hspace{1 cm}
    \begin{subfigure}[t]{0.1\textwidth} 
    \includegraphics[width=\textwidth]{legend.pdf}
    \end{subfigure}
    \caption{Compressive failure of concrete. Cracking patterns for the confined sample, as described by the phase field $\phi$ contours: (a) axisymmetric 2D results, and (b) 3D visualisation. The ratio between the applied pressure and the prescribed displacement equals $P/u_y=10$ MPa/mm.}
    \label{fig:CT-confined}
\end{figure}

Qualitative differences are found between the cracking patterns observed for the confined and unconfined experiments. As shown in Fig. \ref{fig:CT-unconfined}, in the unconfined specimen the crack starts from the edge and propagates gradually towards the centre, creating a cone shape fracture. This is in agreement with the cracking patterns observed experimentally for brittle solids in the absence of confinement \cite{Pollard2010,Hoshhino1972}. However, in the confined specimen, see Fig. \ref{fig:CT-confined}, the crack nucleates at the centre of the sample and then propagates towards the surface, exhibiting a double shear failure mode. Such a cracking pattern has also been reported in experiments conducted under confinement pressures \cite{Hoshhino1972}. Of interest for future work is the analysis of the influence of friction between the sample and the comopression plates, which can be readily be incorporated into the present framework and has been argued to influence cracking patters \cite{Jaeger2009,Freddi2011}.

\subsection{Localised failure of a soil slope}
\label{Sec:Rsoil}

Finally, in our last case study, we compare the predictions of the Drucker-Prager strain energy decomposition formulation to those obtained with what are arguably the most widely use strain energy decompositions in the literature: the volumetric-deviatoric split by Amor \textit{et al.} \cite{Amor2009} and the spectral decomposition by Miehe and co-workers \cite{Miehe2010a}. First, the damaged and stored (elastic) strain energy densities are defined for these two approaches, following the terminology of Section \ref{Sec:Phase field fracture formulation}. Thus, the volumetric-deviatoric split is characterised by,
\begin{equation}
    \psi_d \left( \bm{\varepsilon} \right) = \frac{1}{2} K \langle \text{tr} \left( \bm{\varepsilon} \right) \rangle^2_+ + \mu \left( \bm{\varepsilon}' : \bm{\varepsilon}' \right) \, , \,\,\,\,\,\,\,\,\,\, \psi_s \left( \bm{\varepsilon} \right) = \frac{1}{2} K \langle \text{tr} \left( \bm{\varepsilon} \right) \rangle^2_- \, .
    \label{eq:DevSplit}
\end{equation}

\noindent Here, $\langle a \rangle_{\pm}=\left( a \pm |a| \right)/2$, and $\bm{\varepsilon}'=\bm{\varepsilon}-\text{tr}\left( \bm{\varepsilon}\right) \bm{I}/3$. While the strain energy decomposition by Miehe \textit{et al.} \cite{Miehe2010a} reads,
\begin{equation}
   \psi_d \left( \bm{\varepsilon} \right) =  \frac{1}{2} \lambda \langle \text{tr} \left( \bm{\varepsilon} \right) \rangle^2_+ + \mu \text{tr} \left[ \left( \bm{\varepsilon}^+ \right)^2 \right] \, , \,\,\,\,\,\,\,\,\,\, \psi_s \left( \bm{\varepsilon} \right) =  \frac{1}{2} \lambda \langle \text{tr} \left( \bm{\varepsilon} \right) \rangle^2_- + \mu \text{tr} \left[ \left( \bm{\varepsilon}^- \right)^2 \right] \, ,
   \label{eq:SpectralSplit}
\end{equation}

\noindent where a spectral decomposition is applied to the strain tensor, such that $\bm{\varepsilon}^\pm=\sum_{a=1}^3 \langle \varepsilon_I \rangle_\pm \mathbf{n}_I \otimes \mathbf{n}_I$, with $\varepsilon_I$ and $\mathbf{n}_I$ being, respectively, the strain principal strains and principal strain directions (with $I=1,2,3$).\\

\begin{figure}[H]
    \centering
    \includegraphics[width=0.5\textwidth]{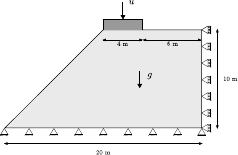}
    \caption{Localised failure of a soil slope. Geometry and boundary conditions.}
    \label{fig:SF-config}
\end{figure}

The boundary value problem under consideration is inspired by the work by Regueiro and Borja \cite{Regueiro2001}, where a strong discontinuity approach was used to predict the stability of a soil slope. This problem was also recently investigated by Fei and Choo \cite{Fei2020} using a phase field-based frictional shear fracture model. The geometry, dimensions and boundary conditions are given in Fig. \ref{fig:SF-config}. A rigid foundation is placed at the crest of the slope, as shown in Fig. \ref{fig:SF-config}. First, a gravity load is applied, followed by a vertical displacement that is prescribed at the centre of the rigid foundation. The material properties of the soil are given by $E=10$ MPa, $\nu=0.4$, $\ell=0.1$ m, $G_c=0.2$ kJ/m$^2$, and $B=0.12$. Approximately 50,000 quadrilateral linear elements are used, with the mesh being refined in the crack propagation region through an iterative process. In all cases, the characteristic size of the elements in the damaged region is five times smaller than the phase field length scale $\ell$. 

\begin{figure}[H]
    \centering
    \begin{subfigure}[t]{0.4\textwidth} 
    \includegraphics[width=\textwidth]{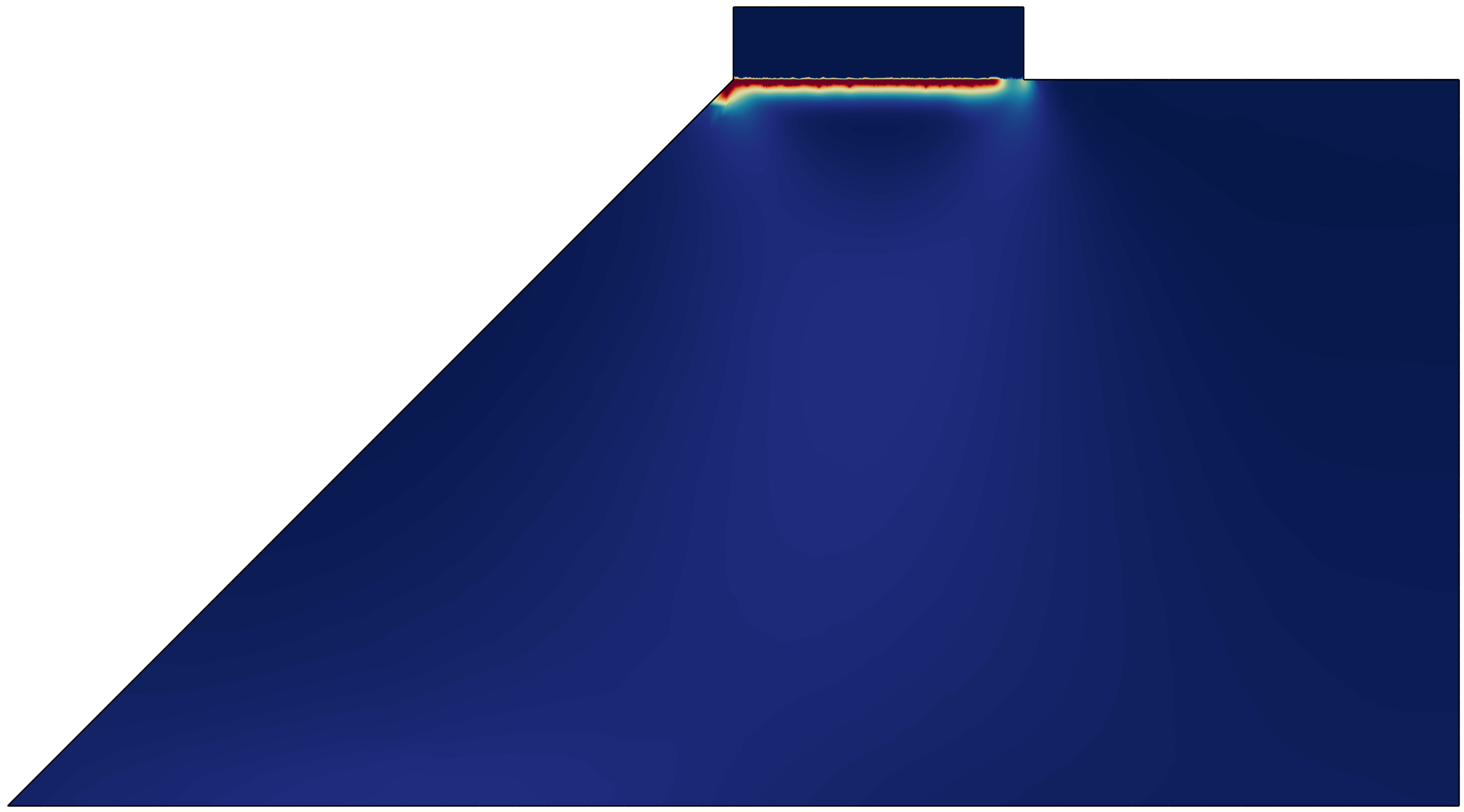} 
    \caption{Volumetric-deviatoric split}
    \label{fig:SF-Phi-a}
    \end{subfigure} \hspace{.5 cm}
    \begin{subfigure}[t]{0.4\textwidth} 
    \includegraphics[width=\textwidth]{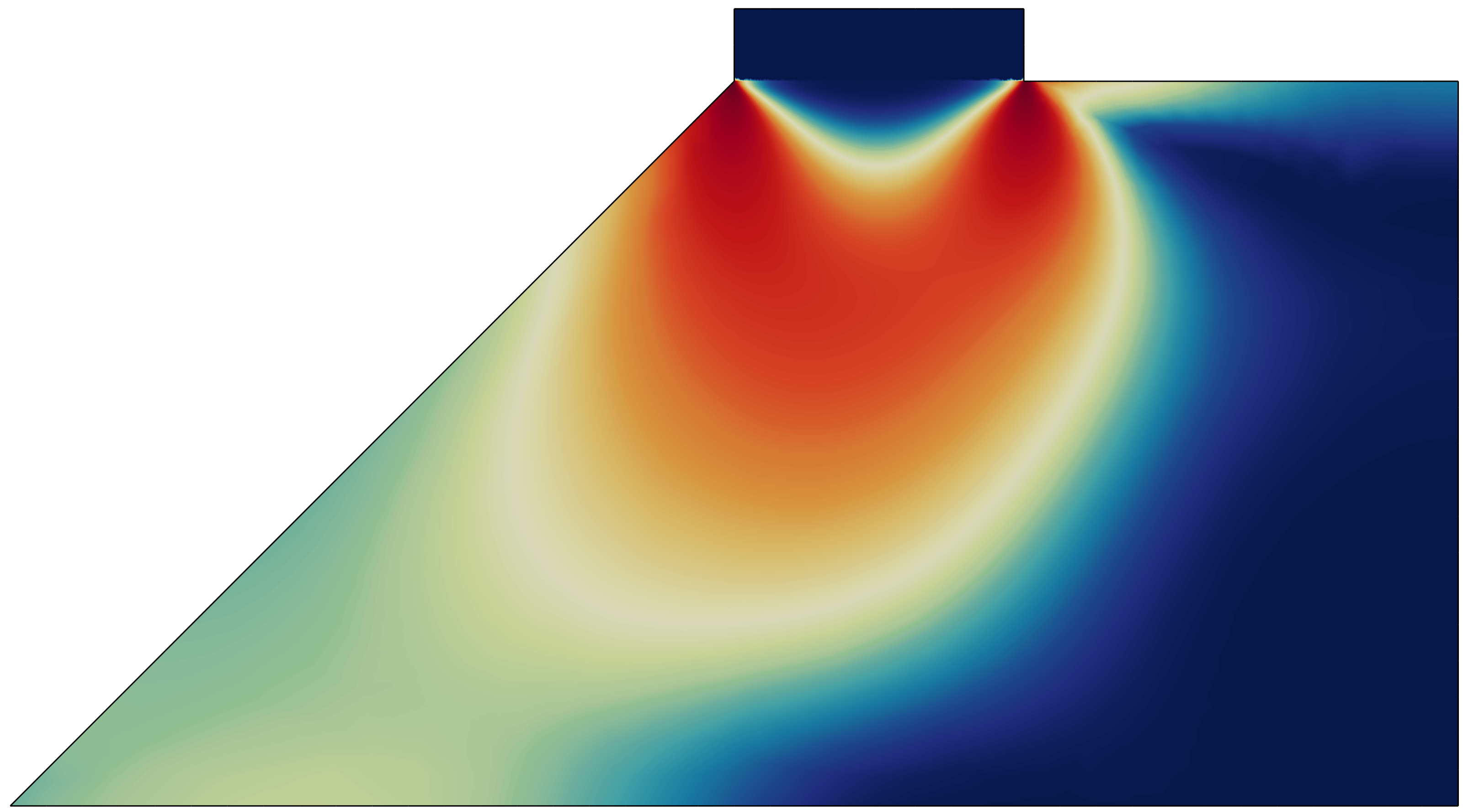} 
    \caption{Spectral decomposition}
    \label{fig:SF-Phi-b}
    \end{subfigure} \hspace{.5 cm}
    \begin{subfigure}[t]{0.4\textwidth} 
    \includegraphics[width=\textwidth]{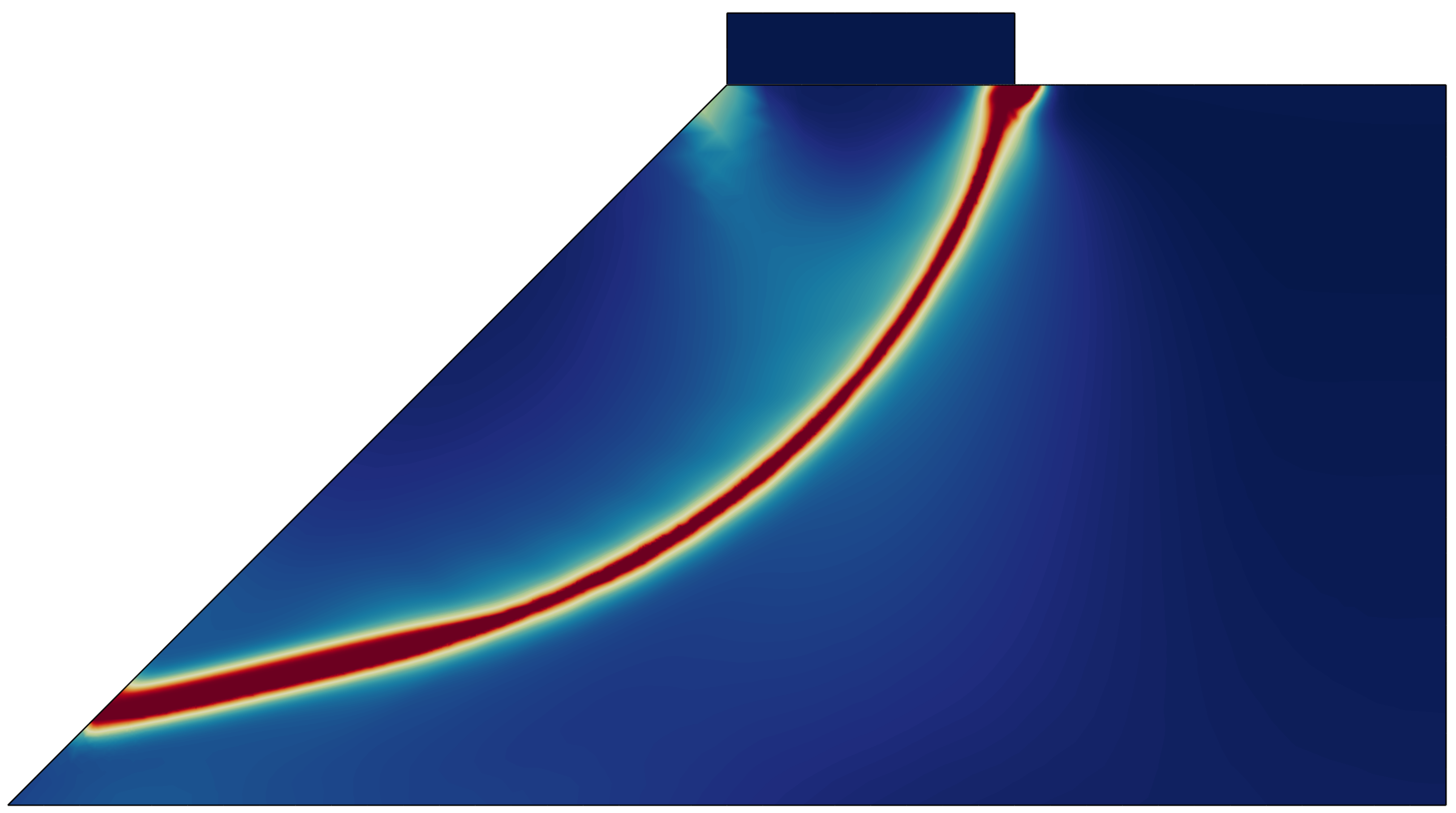}
    \caption{Drucker-Prager based split}
    \label{fig:SF-Phi-c}
    \end{subfigure}\hspace{.5 cm}
    \begin{subfigure}[t]{0.08\textwidth} 
    \includegraphics[width=\textwidth]{legend.pdf}
    \end{subfigure}
    \caption{Localised failure of a soil slope. Failure patterns as described by the contours of the phase field order parameter for: (a) the volumetric-deviatoric split, Eq. (\ref{eq:DevSplit}), (b) the spectral decomposition, Eq. (\ref{eq:SpectralSplit}), and (c) the Drucker-Prager based split presented, Eqs. (\ref{eq:psi_s3regimes})-(\ref{eq:psi_d3regimes}).}
    \label{fig:SF-Phi}
\end{figure}

The results obtained are given in Fig. \ref{fig:SF-Phi}. The cracking patterns are shown for each of the three strain energy decompositions considered, by means of contours of the phase field order parameter $\phi$. As shown in Fig. \ref{fig:SF-Phi}a, the volumetric-deviatoric split by Amor \textit{et al.} \cite{Amor2009} predicts a localised failure under the rigid foundation. The spectral decomposition by Miehe and co-workers \cite{Miehe2010a} is also unable to adequately capture the localised failure of the soil slope. As shown in Fig. \ref{fig:SF-Phi}b, damage accumulates under the rigid foundation, showing a V-type of failure. On the other hand, the Drucker-Prager formulation presented in Section \ref{Sec:General approach} is able to appropriately simulate the localised failure of the soil slope. Cracking initiates from the right corner of the foundation and propagates towards the edge of the slope, in a very similar pattern to that reported by other numerical experiments \cite{Regueiro2001,Fei2020}. % We are missing some justification here, why do we observe such result for each split? (stress/strain mechanics)

\section{Discussion}
\label{Sec:Discussion}

The aim of the present work is to present a general approach to decompose the phase field fracture driving force, the strain energy density, so as to encompass any arbitrary choice of failure criteria. One important motivation for this work lies in the need to enrich the phase field fracture method to go beyond its assumed symmetric tension-compression fracture behaviour to adequately predict crack nucleation and growth in multi-axial stress states. The potential of the general methodology presented is demonstrating by particularising it to the Drucker-Prager failure surface. In doing so, we establish a connection with the recent work by De Lorenzis and Maurini \cite{Lorenzis2021}. De Lorenzis and Maurini \cite{Lorenzis2021} showed analytically that phase field fracture can be generalised to accommodate arbitrary multiaxial failure surfaces and thus faithfully predict crack nucleation without the need to recur to non-variational models. They also chose to particularise their approach to a Drucker-Prager failure surface. Thus, both works reach the same theoretical outcome from different angles. Since our paper also includes a numerical implementation, it complements and extends the work by De Lorenzis and Maurini \cite{Lorenzis2021}, confirming their findings. It is also worth noting that our analysis is not limited to nucleation but also considers the propagation of cracks until failure. To achieve this, it is here assumed that the same surface in the multiaxial stress space characterises the limit of the elastic domain ($\phi > 0$) and the fully damaged state ($\phi=1$). Several numerical experiments are reported to showcase the ability of the model to predict crack nucleation and growth in boundary value problems exhibiting multi-axial loading and mixed-mode fracture conditions. An alternative approach is that proposed by Kumar \textit{et al.} \cite{Kumar2020a}, where an external driving force is defined to recover a Drucker-Prager failure surface. However, this comes at the cost of losing the variatonal consistency.

\section{Conclusions}
\label{Sec:Conclusions}

We have presented a general framework for determining the strain energy decomposition associated with arbitrary choices of constitutive behaviour and failure criterion. This is of importance for phase field fracture modelling as it opens a new avenue for incorporating multi-axial failure surfaces and thus appropriately capturing crack nucleation in a wide range of materials. In particular, this is needed to predict the compressive failure of brittle and quasi-brittle solids such as concrete and geomaterials. Accordingly, we chose to illustrate our framework by particularising it to the case of a Drucker-Prager failure surface. We numerically implemented the resulting formulation for the strain energy decomposition and used it to simulate fracture phenomena in brittle materials. Specifically, the potential of the Drucker-Prager based formulation presented was showcased by addressing four paradigmatic case studies. The behaviour of a single element undergoing shear deformations and vertical pressure was investigated first. The results showed that the model is capable of capturing the role of friction and dilatancy. The magnitude of the shear stresses attained was highest for higher values of the pressure and of Drucker-Prager's parameter $B$. Direct Shear Tests (DST) were subsequently simulated showing a noticeable influence of the applied pressure. The lower the pressure, the more tortuous the crack path and the lower the magnitude of the residual load predicted. Thirdly, the failure of cylindrical samples under uniaxial and triaxial compression was investigated. The results revealed a qualitative impact of the confinement pressure on both the cracking patterns and the force versus displacement response predicted. Cracking predictions appear to agree with experimental observations, shifting from a cone shape fracture to a double shear failure mode with increasing confinement. Finally, we simulated the localised failure of a soil slope using three different strain energy splits: our Drucker-Prager approach and the widely used volumetric-deviatoric \cite{Amor2009} and spectral \cite{Miehe2010a} decompositions. The results show that only the Drucker-Prager based formulation is able to adequately predict the fracture behaviour. Accordingly, the present work: (i) opens a new avenue for incorporating multi-axial failure criteria in phase field fracture modelling, and (ii) demonstrates the potential of Drucker-Prager based phase field formulations for predicting compressive failures in materials exhibiting asymmetric tension-compression fracture behaviour. 

\section{Acknowledgments}
\label{Sec:Acknowledge of funding}

The authors acknowledge financial support from the Ministry of Science, Innovation and Universities of Spain through grant PGC2018-099695-B-I00. E. Mart\'{\i}nez-Pa\~neda was supported by an UKRI Future Leaders Fellowship (grant MR/V024124/1).


%\processdelayedfloats % This is basically to include the figures here, before the appendix

\appendix

\section{The relation of stress and strain invariants}
\label{App:I1J2Proof}

In the following, we shall show how Eq. (\ref{Eq:strain stress relation}) can be derived for any choice of strain energy density in the form of $\psi (I_1 (\boldsymbol\varepsilon),J_2 (\boldsymbol\varepsilon))$. First, let us express the Cauchy stress as:
\begin{equation}\label{Eg:General stress}
   \bm\sigma \left( \bm{\varepsilon} \right) =\frac{\partial \psi (I_1 (\boldsymbol\varepsilon),J_2 (\boldsymbol\varepsilon))}{\partial \bm{\varepsilon}}= \frac{\partial \psi (I_1 (\boldsymbol\varepsilon),J_2 (\boldsymbol\varepsilon))}{\partial I_1 (\boldsymbol\varepsilon)} \frac{\partial I_1 (\boldsymbol\varepsilon)}{\partial \bm{\varepsilon}} + \frac{\partial \psi (I_1 (\boldsymbol\varepsilon),J_2 (\boldsymbol\varepsilon))}{\partial J_2 (\boldsymbol\varepsilon)} \frac{\partial J_2 (\boldsymbol\varepsilon)}{\partial \bm{\varepsilon}} \, .
\end{equation}

The variations of the first two invariants of the strain tensor are written as,
\begin{equation}
    \frac{\partial I_1 (\boldsymbol\varepsilon)}{\partial \bm{\varepsilon}}=\bm{I}, \quad \frac{\partial J_2 (\boldsymbol\varepsilon)}{\partial \bm{\varepsilon}}=\bm{\varepsilon}'
\end{equation}
\noindent where $\bm{I}$ denotes the identity tensor and $\bm{\varepsilon}'$ is the deviatoric part of strain tensor. On the other side, the first invariant of the Cauchy stress tensor is given by
\begin{equation}
     I_1 (\boldsymbol\sigma)=\text{tr}(\boldsymbol\sigma)=\text{tr} \left(\frac{\partial \psi (I_1 (\boldsymbol\varepsilon),J_2 (\boldsymbol\varepsilon))}{\partial \boldsymbol\varepsilon}\right)= \frac{\partial \psi (I_1 (\boldsymbol\varepsilon),J_2 (\boldsymbol\varepsilon))}{\partial I_1 (\boldsymbol\varepsilon)} \text{tr}\left(\frac{\partial I_1 (\boldsymbol\varepsilon)}{\partial \boldsymbol\varepsilon}\right) + \frac{\partial \psi (I_1 (\boldsymbol\varepsilon),J_2 (\boldsymbol\varepsilon))}{\partial J_2 (\boldsymbol\varepsilon)} \text{tr}\left(\frac{\partial J_2 (\boldsymbol\varepsilon)}{\partial \boldsymbol\varepsilon}\right)
    \label{eq:C3}
\end{equation}

Eq. (\ref{eq:C3}) can be simplified by considering $\text{tr}\left( \partial I_1 (\boldsymbol\varepsilon)/\partial \boldsymbol\varepsilon \right)=3$ and $ \text{tr}\left(\partial J_2 (\boldsymbol\varepsilon)/\partial \boldsymbol\varepsilon\right)=0$, such that
\begin{equation}
   I_1 (\boldsymbol\sigma)=3 \frac{\partial \psi (I_1 (\boldsymbol\varepsilon),J_2 (\boldsymbol\varepsilon))}{\partial I_1 (\boldsymbol\varepsilon)}.
    \label{Eg:I1proof}
\end{equation}
\noindent which corresponds to Eq. (\ref{Eq:strain stress relation})a, the equation relating the first invariant of stress $I_1 (\boldsymbol\sigma)$ with the first invariant of strain $I_1 (\boldsymbol\varepsilon)$. Next, we use Eqs. (\ref{Eg:General stress}) and (\ref{Eg:I1proof}) to formulate the deviatoric part of the Cauchy stress tensor $\bm{\sigma}'$ as
\begin{equation}\label{Eq:DevSigma}
   \bm{\sigma}'= \bm{\sigma} - \frac{1}{3} \text{tr} \left( \bm{\sigma}\right) \bm{I}=\bm{\varepsilon}' \frac{\partial \psi (I_1 (\boldsymbol\varepsilon),J_2 (\boldsymbol\varepsilon))}{\partial J_2 (\boldsymbol\varepsilon)}  \, .
\end{equation}
Then, Eq. (\ref{Eq:strain stress relation})b, relating the second stress invariant $J_2 (\boldsymbol\sigma)$ with its strain-based counterpart $J_2 (\boldsymbol\varepsilon)$ can be obtained by substituting Eq. (\ref {Eq:DevSigma}) into the definition of $J_2 (\boldsymbol\sigma)$, rendering
\begin{equation}
    J_2 (\boldsymbol\sigma)=\frac{1}{2} \text{tr}\left((\boldsymbol\sigma')^2\right) =\frac{1}{2} \text{tr}\left((\boldsymbol\varepsilon')^2\right) \left(\frac{\partial \psi (I_1 (\boldsymbol\varepsilon),J_2 (\boldsymbol\varepsilon))}{\partial J_2 (\boldsymbol\varepsilon)}\right)^2=J_2(\boldsymbol\varepsilon) \left(\frac{\partial \psi (I_1 (\boldsymbol\varepsilon),J_2 (\boldsymbol\varepsilon))}{\partial J_2 (\boldsymbol\varepsilon)}\right)^2.
    \label{Eg:J2proof}
\end{equation}

\section{Strain-based mapping of the stress state scenarios}
\label{App:3regimes}

Any relevant stress state can be classified as one of three potential scenarios in the ($I_1 (\boldsymbol\sigma)$,$\sqrt{J_2(\boldsymbol\sigma)}$) stress space. However, for numerical reasons, the stored (reversible) $\psi_s$ and damaged $\psi_d$ strain energy densities are formulated in terms of the strain tensor $\bm{\varepsilon}$, see Eqs. (\ref{eq:psi_s3regimes})-(\ref{eq:psi_d3regimes}). Thus, for completeness, we proceed to describe the derivation of Eqs. (\ref{eq:psi_s3regimes})-(\ref{eq:psi_d3regimes}) for the stress scenarios discussed in Section \ref{Sec:General approach}.\\

Consider first the third regime, given by Eqs. (\ref{eq:psi_s3regimes})c and (\ref{eq:psi_d3regimes})c, where $I_1 (\boldsymbol\sigma)<0$ and the stress state is below the failure envelope. Under these conditions, damage does not evolve and consequently the stored part of the strain energy density equals the total one $\psi_s(\boldsymbol\varepsilon)=\psi_0(\boldsymbol\varepsilon)$. Specifically, the stress state in this regime fulfills the following:
\begin{equation}
    \sqrt{J_2(\boldsymbol\sigma)} < B I_1 (\boldsymbol\sigma) \quad \,\,\, \text{and} \quad \,\,\, I_1   (\boldsymbol\sigma) \leq 0 \, .
    \label{Eq:1st-Con}
\end{equation}

\noindent Where the stress invariants can be written as,
\begin{equation}
\begin{aligned}
&I_1 (\boldsymbol\sigma)=3 \frac{\partial \psi(\boldsymbol\varepsilon)}{\partial I_1(\boldsymbol\varepsilon)}=3 g (\phi) \frac{\partial \psi_0(\boldsymbol\varepsilon)}{\partial I_1(\boldsymbol\varepsilon)}+3 (1-g (\phi)) \frac{\partial \psi_s(\boldsymbol\varepsilon)}{\partial I_1(\boldsymbol\varepsilon)} \, ,\\
&J_2 (\boldsymbol\sigma)=J_2(\boldsymbol\varepsilon) \left(\frac{\partial \psi(\boldsymbol\varepsilon)}{\partial J_2(\boldsymbol\varepsilon)} \right)^2=g (\phi) J_2(\boldsymbol\varepsilon) \left(\frac{\partial \psi_0(\boldsymbol\varepsilon)}{\partial J_2(\boldsymbol\varepsilon)} \right)^2+(1-g (\phi)) J_2(\boldsymbol\varepsilon) \left(\frac{\partial \psi_s(\boldsymbol\varepsilon)}{\partial J_2(\boldsymbol\varepsilon)} \right)^2 \, .
\end{aligned}
\label{Eq:I1J2SE}
\end{equation}

\noindent Considering that, in this scenario, $\psi_s(\boldsymbol\varepsilon) \equiv \psi_0 (\boldsymbol\varepsilon)$ and inserting Eq. (\ref{Eq:I1J2SE}) into the first condition of Eq. (\ref{Eq:1st-Con}), one reaches
\begin{equation}
     \sqrt{J_2(\boldsymbol\varepsilon)} \frac{\partial \psi_0(\boldsymbol\varepsilon)}{\partial J_2(\boldsymbol\varepsilon)} < 3 B   \frac{\partial \psi_0(\boldsymbol\varepsilon)}{\partial I_1(\boldsymbol\varepsilon)}
     \label{Eq:1st1st-con}
\end{equation}

\noindent Now, recalling the definition of $\psi_0$, Eq. (\ref{Eq:ElasticStrainEnergy}), Eq. (\ref{Eq:1st1st-con}) can be re-formulated as
\begin{equation}
2 \mu \sqrt{J_2 (\boldsymbol\varepsilon)} < 3 B K I_1 (\boldsymbol\varepsilon)
     \label{Eq:1st1st-con2}
\end{equation}

\noindent On the other side, the second condition of Eq. (\ref{Eq:1st-Con}) can be described as a function of the strain tensor as follows,
\begin{equation}
     3 \frac{\partial \psi_0(\boldsymbol\varepsilon)}{\partial I_1(\boldsymbol\varepsilon)} \leq 0 
    \label{Eq:1st2nd-con}
\end{equation}

\noindent Implying that $I_1(\boldsymbol\varepsilon) \leq 0$. However, this has already been satisfied by Eq. (\ref{Eq:1st1st-con2}) as $\sqrt{J_2 (\boldsymbol\varepsilon)}$ is a positive value and the parameter $B$ is always zero or negative, such that $I_1(\boldsymbol\varepsilon)$ must be negative to satisfy Eq. (\ref{Eq:1st1st-con}).\\

The second regime in the ($I_1 (\boldsymbol\sigma)$,$\sqrt{J_2(\boldsymbol\sigma)}$) stress space corresponds to that where $I_1 (\boldsymbol\sigma) \leq 0$ and the stress state is above the failure criterion; i.e.,
\begin{equation}
    \sqrt{J_2(\boldsymbol\sigma)} \geq B I_1 (\boldsymbol\sigma) \quad \,\,\, \text{and} \quad \,\,\, I_1   (\boldsymbol\sigma) \leq 0 \, .
     \label{Eq:1st-ConB}
\end{equation}

\noindent Given that Eq. (\ref{Eq:1st1st-con}) provides the strain condition for the case where the stress state is below the failure criterion, it follows that the relevant condition for the second regime where the stress state is above the failure criterion is given by 
\begin{equation}
    2 \mu \sqrt{J_2 (\boldsymbol\varepsilon)} \geq 3 B I_1 (\boldsymbol\varepsilon)
    \label{eq:Acond1_reg2}
\end{equation}

\noindent Then, the second condition in Eq. (\ref{Eq:1st-ConB}) can be expressed as:
\begin{equation}
    g(\phi) K I_1(\boldsymbol\varepsilon)+  \frac{K \mu}{9 B^2 K + \mu} (1-g(\phi)) \left(I_1(\boldsymbol\varepsilon)+6 B \sqrt{J_2(\boldsymbol\varepsilon)} \right) \leq 0  \, .
\end{equation}

\noindent Which, considering that $g(\phi=1)=0$, can be reduced to,
\begin{equation}
     I_1(\boldsymbol\varepsilon) \leq -6 B \sqrt{J_2(\boldsymbol\varepsilon)}
     \label{eq:Acond2_reg2}
\end{equation}

\noindent Accordingly, the conditions for the second regime, in terms of the strain tensor, are given by (\ref{eq:Acond1_reg2}) and (\ref{eq:Acond2_reg2}).\\ 

The remaining conditions are applicable for the first regime in the stress space, where $I_1 (\boldsymbol\sigma)$ is positive:
\begin{equation}
\mu \sqrt{J_2 (\boldsymbol\varepsilon)} \geq 3 B K I_1 (\boldsymbol\varepsilon) \, ; \,\,\,\,\,\, \,\,\,\, \quad -6 B \sqrt{J_2(\boldsymbol\varepsilon)} < I_1(\boldsymbol\varepsilon) \, ,
\label{Eq:3rd-con}
\end{equation}

\noindent where the first condition can be neglected as it is satisfied by the second one.

\section{Additional details of the finite element implementation}
\label{App:FEM}

\subsection{Strong and weak formulations}
\label{App:FEMstrongweak}

Considering Eq. (\ref{eq:E_ell}) and the constitutive choices in Eq. (\ref{eq:ConsPF}), Griffith's regularised energy functional can be formulated as,
\begin{equation}
    \mathcal{E}_\ell  = \int_\Omega \psi_s \left( \bm{\varepsilon} \left( \mathbf{u} \right) \right) + \left( 1 - \phi \right)^2 \psi_d \left( \bm{\varepsilon} \left( \mathbf{u} \right) \right)  \, \text{d} V + \int_{V} G_c \left(\frac{1}{2\ell}\phi^2 + \frac{\ell}{2} \lvert\nabla \phi\rvert^{2}\right) \, \text{d} V  
\end{equation}

The stationary of $\mathcal{E}_\ell$ with respect to the primal kinematic variables renders,
\begin{equation}
    \partial \mathcal{E}_\ell  = \int_{\Omega}\left\{\left[\left( 1 - \phi \right)^2 \frac{\partial \psi_d \left( \bm{\varepsilon} \right)}{\partial \bm{\varepsilon}} + \frac{\partial \psi_s \left( \bm{\varepsilon} \right)}{\partial \bm{\varepsilon}} \right] \delta \boldsymbol{\varepsilon}-2(1-\phi) \delta \phi \psi_{d}(\boldsymbol{\varepsilon})+G_{c}\left[\frac{1}{\ell} \phi \delta \phi+\ell \nabla \phi \cdot \nabla \delta \phi\right]\right\} \mathrm{d} V
\end{equation}

Accordingly, the strong form can be readily derived by considering the variation in the external work,
\begin{equation}
\delta W_{e x t}=\int_{\Omega} \mathbf{b} \cdot \delta \mathbf{u} \mathrm{d} V+\int_{\partial \Omega_{h}} \mathbf{h} \cdot \delta \mathbf{u} \mathrm{d} A
\end{equation}

\noindent enforcing equilibrium of the external and internal virtual works,
\begin{equation}
\partial \mathcal{E}_\ell -\delta W_{e x t}=0
\end{equation}

\noindent and making use of Gauss' divergence theorem,
\begin{align}
\nabla \cdot \left[\left( 1 - \phi \right)^2 \frac{\partial \psi_d \left( \bm{\varepsilon} \right)}{\partial \bm{\varepsilon}} + \frac{\partial \psi_s \left( \bm{\varepsilon} \right)}{\partial \bm{\varepsilon}} \right] + \mathbf{b} &= \boldsymbol{0}   \hspace{3mm} \rm{in}  \hspace{3mm} \Omega \nonumber \\ 
G_{c}  \left( \dfrac{\phi}{\ell}  - \ell \nabla^2 \phi \right) -2(1-\phi) \, \psi_d  &= 0 \hspace{3mm} \rm{in} \hspace{3mm} \Omega 
\label{eq:AppB_balance}
\end{align}

% We should add BCs here.

\subsection{Heat transfer analogy}

As discussed in Refs. \cite{Materials2021,AES2021}, we exploit the analogy with heat transfer to facilitate the numerical implementation of the phase field evolution equation. In the presence of a heat source $r$, the steady state equation for heat transfer has the following form,
\begin{equation}\label{Eq:HeatTransfer}
k \nabla^{2} T=-r
\end{equation}

\noindent where $T$ is the temperature, and $k$ is the thermal conductivity. Eq. (\ref{Eq:HeatTransfer}) is analogous to the phase field evolution equation (\ref{eq:AppB_balance}b) upon assuming $T \equiv \phi$, $k=1$, and defining the heat source $r$ as follows:
\begin{equation}
r=\frac{2(1-\phi) \mathcal{H}}{\ell G_{c}}-\frac{\phi}{\ell^{2}}
\end{equation}

\noindent where, as discussed in Section \ref{Sec:General approach}, $\mathcal{H}=\text{max} \, \psi_d (t)$ is a history field introduced to enforce damage irreversibility. Finally, the variation of the heat source with respect to the phase field (temperature) is derived as,
\begin{equation}
\frac{\partial r}{\partial \phi}=-\frac{2 \mathcal{H}}{\ell G_{c}}-\frac{1}{\ell^{2}}
\end{equation}

\subsection{Finite element discretisation}

By exploiting the heat transfer analogy, one can implement the phase field formulation described in this paper into the finite element package \texttt{ABAQUS} using only a user material subroutine (\texttt{UMAT}). I.e., there is no need to explicitly define and implement the element stiffness matrix $\bm{K}^e$ and the element residual vector $\bm{R}^e$. However, these are derived here for completeness. Consider the equilibrium of the external and internal virtual works presented in Section \ref{App:FEMstrongweak}. Decoupling the displacement and phase field problems, the weak form equations read,
\begin{equation}
  \int_\Omega \Big\{ \left[\left( 1 - \phi \right)^2 \frac{\partial \psi_d \left( \bm{\varepsilon} \right)}{\partial \bm{\varepsilon}} + \frac{\partial \psi_s \left( \bm{\varepsilon} \right)}{\partial \bm{\varepsilon}} \right] : \delta \bm{\varepsilon} -\mathbf{b} \cdot \delta \mathbf{u}  \Big\} \text{d} V - \int_{\partial \Omega_{h}} \mathbf{h} \cdot \delta \mathbf{u} \mathrm{d} A=0  \, .
\end{equation}
\begin{equation}
\int_{\Omega} \left\{ {- 2 (1 - \phi)\delta \phi} \, \mathcal{H} +
        G_c \left[ \frac{1}{ \ell}\phi  \delta \phi + \ell \nabla \phi  \nabla \delta \phi \right] \right\}  \, \mathrm{d}V = 0  \, .
\end{equation}

Now, consider the following finite element discretisation. Adopting Voig notation, the nodal variables for the displacement field $\mathbf{\hat{u}}$, and the phase field $\hat{\phi}$ are interpolated as: 
\begin{equation}\label{eq:Ndiscret}
\mathbf{u} = \sum_{i=1}^m \bm{N}_i \hat{\mathbf{u}}_i, \hspace{1cm} \phi =  \sum_{i=1}^m N_i \hat{\phi}_i \, ,
\end{equation}
\noindent where $N_i$ is the shape function associated with node $i$ and $\bm{N}_i$ is the shape function matrix, a diagonal matrix with $N_i$ in the diagonal terms. Also, $m$ is the total number of nodes per element and $\hat{\mathbf{u}}_i$ and $\hat{\phi}_i$ respectively denote the displacement and phase field at node $i$. In a similar manner, the associated gradient quantities can be discretised using the corresponding \textbf{B}-matrices, containing the derivative of the shape functions, such that:
\begin{equation}\label{eq:Bdiscret}
\bm{\varepsilon} = \sum\limits_{i=1}^m \bm{B}^{\bm{u}}_i \hat{\mathbf{u}}_i, \hspace{0.8cm}  \nabla\phi =  \sum\limits_{i=1}^m \mathbf{B}_i \hat{\phi}_i \, .
\end{equation}

The discretised residuals for each primal kinematic variable are then given by:
\begin{align}
    & \mathbf{R}_i^\mathbf{u} = \int_\Omega \left\{\left( 1 - \phi \right)^2 \left(\bm{B}^\mathbf{u}_i\right)^T \frac{\partial \psi_d \left( \bm{\varepsilon} \right)}{\partial \bm{\varepsilon}} + \left(\bm{B}^\mathbf{u}_i\right)^T \frac{\partial \psi_s \left( \bm{\varepsilon} \right)}{\partial \bm{\varepsilon}} \right\} \, \text{d}V \, -\int_{\Omega}\left(\mathbf{N}_{i}^{\mathrm{u}}\right)^{T} \mathbf{b} \mathrm{d} V-\int_{\partial \Omega_{h}}\left(\mathbf{N}_{i}^{\mathrm{u}}\right)^{T} \mathbf{h} \mathrm{d} A , \\
    & 
\mathbf{R}_{i}^{\phi}=\int_{\Omega}\left\{-2(1-\phi) N_{i} \mathcal{H}+G_{c}\left[\frac{1}{\ell} N_{i} \phi+\ell\left(\mathbf{B}_{i}^{\phi}\right)^{T} \nabla \phi\right]\right\} \mathrm{d} V
\end{align}

And the consistent tangent stiffness matrices $\bm{K}$ are obtained by differentiating the residuals with respect to the incremental nodal variables:
\begin{align}
    & \bm{K}_{ij}^{\mathbf{u}} = \frac{\partial \bm{R}_{i}^{\bm{u}} }{\partial \bm{u}_{j} } =         \int_{\Omega} \left\{ ( 1 - \phi)^2 {(\bm{B}_{i}^{\bm{u}})}^{T} \bm{C}_d \, \bm{B}_{j}^{\bm{u}} + {(\bm{B}_{i}^{\bm{u}})}^{T} \bm{C}_s \, \bm{B}_{j}^{\bm{u}}  \right\} \, \text{d}V \, , \\
    & \bm{K}_{ij}^{\phi} = \frac{\partial R_{i}^{\phi} }{ \partial \phi_{j} } =  \int_{\Omega} \left\{ \left( 2 \mathcal{H} + \frac{G_{c}}{ \ell} \right) N_{i} N_{j} +G_{c} \ell   \, \mathbf{B}_i^T\mathbf{B}_j \right\} \, \text{d}V \, ,
\end{align}

\noindent Here, the material jacobian $\bm{C}_s$ can be defined as:
\begin{equation}
\bm{C}_s=\frac{\partial \psi_s}{\partial \boldsymbol\varepsilon \partial \boldsymbol\varepsilon}=
\begin{cases}
0 & \text{for} \quad -6B\sqrt{J_2 (\boldsymbol\varepsilon)} < I_1 (\boldsymbol\varepsilon) \\
\bm{C}_s^{DP} & \text{for} \quad -6B\sqrt{J_2 (\boldsymbol\varepsilon)} \geq I_1 (\boldsymbol\varepsilon) \,\, \& \,\,  2 \mu \sqrt{J_2 (\boldsymbol\varepsilon)} \geq 3 B K I_1 (\boldsymbol\varepsilon) \\
\bm{C}_0 & \text{for} \quad 2 \mu \sqrt{J_2 (\boldsymbol\varepsilon)} < 3 B K I_1 (\boldsymbol\varepsilon) 
\end{cases}
\end{equation}

\noindent where $\bm{C}_0$ is undamaged elastic tangent stiffness and $\bm{C}_s^{DP}$ can be written as:

\begin{equation}
\begin{aligned}
(C_s^{DP})_{ijkl}=\frac{K \mu}{9 B^2 K + \mu}\left(\frac{\partial I_{1}}{\partial \varepsilon_{i j}}+\frac{3 B}{\sqrt{J_{2}}} \frac{\partial J_{2}}{\partial \varepsilon_{i j}}\right)\left(\frac{\partial I_{1}}{\partial \varepsilon_{k l}}+\frac{3 B}{\sqrt{J_{2}}} \frac{\partial J_{2}}{\partial \varepsilon_{K l}}\right)+ \\ \left(\frac{6 B a_{1}\left(I_{1}+6 B \sqrt{J_{2}}\right)}{\sqrt{J_{2}}}\right)\left(\frac{\partial^{2} J_{2}}{\partial \varepsilon_{i j} \partial \varepsilon_{k l}}-\frac{1}{2 J_{2}} \frac{\partial J_{2}}{\partial \varepsilon_{i j}} \frac{\partial J_{2}}{\partial \varepsilon_{k l}}\right)
\end{aligned}
\end{equation}

Finally, $\bm{C}_d$ is obtained by exploiting the fact that $\psi_d=\psi_0-\psi_s$:
\begin{equation}
\bm{C}_d=\frac{\partial \psi_d}{\partial \boldsymbol\varepsilon \partial \boldsymbol\varepsilon}=\frac{\partial \psi_0}{\partial \boldsymbol\varepsilon \partial \boldsymbol\varepsilon}-\frac{\partial \psi_s}{\partial \boldsymbol\varepsilon \partial \boldsymbol\varepsilon}=\bm{C}_0-\bm{C}_s
\end{equation}

\bibliographystyle{elsarticle-num} 
\bibliography{library}


\end{document}